\documentclass[a4paper,11pt]{article}
\usepackage{jheppub} 
\usepackage{lineno}
\usepackage{graphicx}
\usepackage{bm}
\usepackage{mathrsfs}
\usepackage{amsthm,amsmath,amssymb}
\usepackage{verbatim}
\usepackage{siunitx} 
\usepackage{booktabs} 
\usepackage{orcidlink}
\usepackage{hyperref}

\arxivnumber{2606.08139} 

\title{Retarded Correlators of Charge Transport in a Magnetic Field}

\author[a]{Xuan Zhao\,\orcidlink{0009-0007-6093-5464},}
\author[a]{Qiuze Sun\,\orcidlink{0009-0005-5371-8029},}
\author[a]{Yi Wang\,\orcidlink{0009-0004-4811-2391},}
\author[b]{Jin Hu\,\orcidlink{0000-0003-1179-4603}}

\affiliation[a]{Department of Physics, Tsinghua University, Beijing 100084, China}
\affiliation[b]{Department of Physics, Fuzhou University, Fujian 350116, China}

\emailAdd{zhaox21@mails.tsinghua.edu.cn}
\emailAdd{sunqz21@mails.tsinghua.edu.cn}
\emailAdd{y-wang23@mails.tsinghua.edu.cn}
\emailAdd{hu-j23@fzu.edu.cn}

\abstract{We study charge transport in a magnetized relativistic plasma using kinetic theory within the relaxation-time approximation. By exactly solving the linearized Boltzmann equation in a uniform magnetic field, we obtain an analytic solution for the distribution function in terms of Bessel functions. Using this solution, we compute the full set of retarded current-current correlators and verify the Ward identities. In the hydrodynamic limit, we extract the charge diffusion modes, demonstrating that the transverse diffusion coefficient is strongly suppressed by the magnetic field, scaling as $1/B_0^2$ in the strong-field regime, while the longitudinal diffusion remains unaffected. Furthermore, we analyze the non-hydrodynamic branch cuts in the complex frequency plane, determining their kinematic thresholds and identifying the underlying wave-particle interactions as longitudinal Landau damping and transverse cyclotron damping.}

\keywords{Finite Temperature or Finite Density, Quark-Gluon Plasma}

\begin{document}
\maketitle
\flushbottom

%%%%%%%%%%%%%%%%%%%%%%%%%%%%%%%%%%%%%%%%%%%%%%
\section{Introduction}
%%%%%%%%%%%%%%%%%%%%%%%%%%%%%%%%%%%%%%%%%%%%%%
Relativistic kinetic theory provides a framework for describing the non-equilibrium dynamics and transport phenomena of relativistic many-particle systems. By systematically connecting the evolution of single-particle distribution functions to macroscopic dynamics of conserved currents and collective excitations, the relativistic Boltzmann equation has become a standard tool for modeling the space-time evolution of relativistic matter~\cite{groot1980relativistic}. It has been extensively applied to the study of the quark-gluon plasma (QGP) produced in ultra-relativistic heavy-ion collisions~\cite{Arnold:2002zm, Arnold:2003zc, Kurkela:2011ti, Wang:2021oqq, Lenkiewicz:2019glw, Xu:2014ega,Hu:2022vph}, as well as dense matter in compact stars~\cite{Schmitt:2017efp, Shternin:2008es, Shternin:2022fii} and the primordial plasma of the early universe~\cite{bernstein1988kinetic, kolb1994early}. Compared to purely macroscopic effective descriptions such as hydrodynamics, kinetic theory retains direct access to the underlying phase-space distribution as long as the quasiparticle picture is allowed, and provides a  description of the dynamical evolution from the collisionless free-streaming regime at early times to the viscous hydrodynamic regime approaching local thermal equilibrium~\cite{Berges:2020fwq, Kurkela:2018vqr}.

To characterize the collective excitations and transport properties of many-body systems, retarded correlators serve as a central tool. Within linear response theory, these correlators encode the causal response of a system to external perturbations~\cite{Kadanoff:1963axw, kapusta2006finite, Kovtun:2012rj, Amoretti:2020mkp, Amoretti:2022acb}. Crucially, the singularity structure of retarded correlators in the complex frequency plane plays a key role in determining the dynamical evolution of the system. Hydrodynamic poles correspond to long-lived collective excitations such as charge diffusion and sound modes, which govern the late-time approach to equilibrium. By contrast, non-hydrodynamic singularities, which may appear as poles or branch cuts, are associated with microscopic fast degrees of freedom. These modes set the characteristic hydrodynamization time and determine the regime of validity of hydrodynamic descriptions. Motivated by the physical significance of these singularities, extensive research has been devoted to characterizing the analytic structure of retarded correlators. In the context of large $N$ thermal $\mathcal{N}=4$ super Yang-Mills theory at large t'Hooft coupling, studies have shown that the associated correlators exhibit only poles or quasinormal modes~\cite{Hartnoll:2005ju, Kovtun:2005ev, Grozdanov:2016vgg}. In weakly coupled regimes, however, the non-analytic structures become more diverse. Within the framework of kinetic theory, calculations employing the relaxation time approximation (RTA) have revealed the interplay between hydrodynamic poles and the branch cuts associated with non-hydrodynamic degrees of freedom on the principal Riemann sheet~\cite{Romatschke:2015gic, Bajec:2024jez}. Meanwhile, investigations utilizing momentum-dependent RTA  models~\cite{Kurkela:2017xis, Brants:2024wrx}, along with other analytical and numerical simulations~\cite{Gavassino:2024rck, Moore:2018mma, Ochsenfeld:2023wxz}, have shown that branch cuts can also  emerge prominently, i.e., these branch cuts can approach the origin arbitrarily closely and become gapless. In this case, the non-hydrodynamic degrees of freedom could become parametrically long-lived, and the basic picture in which non-hydrodynamic modes are damped out, causing hydrodynamic modes to dominate and hydrodynamic behavior to emerge, breaks down because there is no longer a clear separation between the two. Nevertheless, whether these two sectors remain well separated or not depends on the details of the microscopic interactions: the nature of the interparticle forces (e.g., hard versus soft scattering processes) determines the spectral properties of the linearized collision operator. Therefore, when analyzing the interplay of non-analytic structures in correlation functions, one must perform a case-by-case discussion \cite{Hu:2024tnn}. 

Despite these comprehensive studies of thermal correlators in systems without a background field, a systematic analytical treatment of retarded correlation functions in the presence of a background magnetic field has not yet been fully developed within kinetic theory. Understanding such magnetized dynamics is crucial for  a range of physical environments, ranging from the primordial plasma of the early universe and dense neutron stars to the QGP created in ultra-relativistic heavy-ion collisions. In the latter context, non-central collisions can generate transient magnetic fields on the order of $eB_0\sim m_\pi^2$ or larger~\cite{Skokov:2009qp, Deng:2012pc, Bzdak:2011yy, McLerran:2013hla}. Such background fields break spatial isotropy and modify the transport properties, collective excitations, and nonequilibrium evolution of strongly interacting matter~\cite{Abbasi:2015nka, Hattori:2017usa, Fang:2024sym, Fang:2024skm, Fang:2024hxa, Wang:2021oqq, Guo:2017dzf, Huang:2020wrr,Hu:2022ofv}, leading to anomalous phenomena such as the chiral magnetic effect~\cite{Fukushima:2008xe, Kharzeev:2015znc, Hattori:2016emy, Huang:2015oca, Zhao:2019hta}. For these reasons, elucidating the  response behavior of magnetized relativistic plasmas remains a pressing theoretical priority, and is also of profound physical significance. In this work, we  attempt to close this gap.

In this work, we focus on the charge transport and  analytically compute the complete set of retarded current-current correlators for massive particles in the presence of  a uniform background magnetic field. We begin in Section \ref{sec2 Kinetic theory in the RTA} by introducing the kinetic framework within the relaxation-time approximation and setting up the linearized Boltzmann equation. By employing a Jacobi-Anger expansion, Section \ref{sec3 Exact Solution} derives the exact analytic distribution function in terms of Bessel functions. Utilizing this exact solution, Section \ref{sec4 Correlators} calculates the full set of retarded correlators and explicitly verifies the corresponding Ward identities. Subsequently, Section \ref{sec5 Analytic Structure} delves into the analytic structure of these correlators in the complex frequency plane.  There, we extract the hydrodynamic diffusion mode and reveal that the non-hydrodynamic branch cuts arise from  transverse cyclotron damping and longitudinal Landau damping. Finally, Section \ref{sec6 Summary and Conclusion} summarizes our findings and offers a brief outlook. Further technical details and derivations are relegated to the appendices.

\vspace{0.3cm}
\noindent\emph{Notations and conventions.}---Throughout this paper, we adopt natural units $\hbar = c = k_B = 1$. The Minkowski metric tensor is chosen with the signature $g^{\mu\nu} = \text{diag}(1, -1, -1, -1)$. Greek indices $\mu, \nu$ run over the four spacetime dimensions $(0, 1, 2, 3)$, while Latin indices $i, j$ denote the spatial components $(1, 2, 3)$. Unless otherwise specified, we consider relativistic massive particles satisfying the on-shell condition $p^\mu p_\mu=m^2$, or equivalently, $p^0=\sqrt{\mathbf p^2+m^2}$. The $U(1)$ charge of the particles is denoted by $q$, and we set $q=1$ throughout this work for notational simplicity.

%%%%%%%%%%%%%%%%%%%%%%%%%%%%%%%%%%%%%%%%%%%%%%
\section{Kinetic Theory in the Relaxation Time Approximation}
\label{sec2 Kinetic theory in the RTA}
%%%%%%%%%%%%%%%%%%%%%%%%%%%%%%%%%%%%%%%%%%%%%%
The dynamical evolution of the on-shell single-particle distribution function $f(t,\mathbf{x},\mathbf{p})$ in a relativistic plasma is governed by the relativistic  Boltzmann equation. To model the collision kernel while preserving analytical tractability, we employ the relaxation time approximation (RTA) with a constant relaxation time~\cite{Anderson:1974nyl}. In the presence of an external electromagnetic field, the transport equation takes the form
\begin{equation}
p^\mu\partial_\mu f+F^{i\beta}p_\beta\frac{\partial}{\partial p^i}f=-\frac{p_\mu u^\mu}{\tau_R}\left(f-f_{\mathrm{eq}}\right).
\end{equation}
Here, $F^{\alpha\beta}=\partial^\alpha A^\beta-\partial^\beta A^\alpha$ denotes the electromagnetic field strength tensor, and $\tau_R$ is the phenomenological relaxation time. The local equilibrium distribution $f_{\mathrm{eq}}$ depends on the local temperature $T(x)$, the local chemical potential $\mu(x)$, and the flow velocity $u^\mu(x)$ with $u^\mu u_\mu=1$. The linearized Boltzmann collision operator contains zero modes associated with conservation laws, as well as nonzero modes describing relaxation toward local equilibrium. After the relevant conservation law is enforced through the matching condition, the RTA characterizes the collisional relaxation of the remaining nonconserved sector by a single time scale $\tau_R$. The RTA thus qualitatively captures the essential physical effect of collisions, namely the relaxation of nonequilibrium deviations toward local equilibrium, but does not retain the detailed momentum dependence or the distinct microscopic relaxation rates encoded in the underlying collision operator. Despite these limitations, the RTA provides an insightful and instructive
framework for the analytical study of transport phenomena within kinetic theory.

Below we consider a background state in global thermal equilibrium with constant temperature $T_0$ and chemical potential $\mu_0$ under a constant and uniform magnetic field $\mathbf{B}_0$. Without loss of generality, we align the magnetic field along the $x$-axis. The corresponding background distribution function is given by
\begin{equation}
    f_0(\mathbf{p})=e^{\frac{\mu_0-p_\mu u_0^\mu}{T_0}},
\end{equation}
which is the classical Maxwell--Boltzmann distribution.
Accordingly, quantum-statistical effects, such as Pauli blocking and
Bose enhancement, are not included in the present analysis. This
distribution function satisfies
\begin{equation}
    p^\mu\partial_\mu f_0+F_0^{i\beta}p_\beta\frac{\partial}{\partial p^i}f_0=0,
\end{equation}
with flow velocity $u_0^\mu=(1,0,0,0)$ and electromagnetic field strength tensor
\begin{equation}F_0^{\alpha\beta}=
\begin{pmatrix}
0 & 0 & 0 & 0 \\
0 & 0 & 0 & 0 \\
0 & 0 & 0 & -B_0 \\
0 & 0 & B_0 & 0
\end{pmatrix}.\end{equation}
A detailed derivation of this global equilibrium state can be found in Appendix~\ref{app:equilibrium_gauge}.

Our primary objective is to extract the linear response of this magnetized relativistic plasma and focus on the charge transport. To this end, we introduce a small external $U(1)$ gauge perturbation $\delta A^\mu(t,\mathbf{x})$. To isolate the charge transport channel, we perturb the chemical potential  while keeping the temperature and the flow velocity fixed at their background values. The reason is that turning on the source $\delta A^\mu(t,\mathbf{x})$ naturally induces fluctuations in the chemical potential, as shown in Eq. \eqref{mu}. In contrast, this source  is not directly tied to the temperature or velocity fields. In more general and realistic physical setups, one must additionally turn on the metric source; this then introduces perturbations in both the temperature and velocity fields, thereby enabling the study of coupled transport of energy-momentum and charge. 
  
  The macroscopic thermodynamic variables and the flow velocity are
\begin{align}
    \begin{aligned}
        \mu(t,\mathbf{x})&=\mu_0+\delta\mu(t,\mathbf{x}), \\
        T&=T_0=\mathrm{Const},  \\
        u^\mu&=u_0^\mu=(1,0,0,0).  \\
    \end{aligned}
\end{align}
Correspondingly, the single-particle distribution function $f(t,\mathbf{x},\mathbf{p})$ and the local equilibrium distribution $f_{\mathrm{eq}}(t,\mathbf{x},\mathbf{p})$ are decomposed into the background and a linear fluctuation
\begin{align}
    \begin{aligned}
        f(t,\mathbf{x},\mathbf{p})&=f_0(\mathbf{p})+\delta f(t,\mathbf{x},\mathbf{p}), \\
        f_{\mathrm{eq}}(t,\mathbf{x},\mathbf{p})&=f_0(\mathbf{p})+\delta f_{\mathrm{eq}}(t,\mathbf{x},\mathbf{p}), \\
        \delta f_{\mathrm{eq}}(t,\mathbf{x},\mathbf{p})&=\frac{f_0(\mathbf{p})}{T_0}\delta \mu(t,\mathbf{x}).
    \end{aligned}
\end{align}
Small perturbations $\delta f(t,\mathbf{x},\mathbf{p})$ are required to obey
\begin{equation}
p^\mu\partial_\mu (f_0+\delta f)+(F^{i\beta}_0+\delta F^{i\beta})p_\beta\frac{\partial}{\partial p^i}(f_0+\delta f)=-\frac{p^0}{\tau_R}\left(\delta f-\delta f_{\mathrm{eq}}\right).  \label{RTA equation}
\end{equation}
To solve this differential equation analytically, it is convenient to perform a Fourier transform from spacetime $(t,\mathbf{x})$ to frequency-momentum space $(\omega,\mathbf{k})$.\footnote{Here, we define the Fourier transform and its inverse as follows:\begin{equation*}
    \tilde{Q}(\omega,\mathbf{k})=\int dt d^3\mathbf{x}e^{i(\omega t-\mathbf{k}\cdot\mathbf{x})}{Q}(t,\mathbf{x}),
\end{equation*}
\begin{equation*}
    {Q}(t,\mathbf{x})=\int \frac{d\omega d^3\mathbf{k}}{(2\pi)^4}e^{-i(\omega t-\mathbf{k}\cdot\mathbf{x})}\tilde{Q}(\omega,\mathbf{k}).
\end{equation*}} With this Fourier convention, the electric-field perturbation induced by the gauge-field perturbation $\delta A^\mu$ is
\begin{equation}
    \delta\tilde{\mathbf E}(\omega,\mathbf{k})
    =
    i\left[
        \omega\,\delta\tilde{\mathbf A}(\omega,\mathbf{k})
        -
        \mathbf{k}\,\delta\tilde A^0(\omega,\mathbf{k})
    \right].
\end{equation}
In the following, for notational simplicity, we denote
$\delta\tilde{\mathbf E}$ simply by $\tilde{\mathbf E}$.

By linearizing Eq.~\eqref{RTA equation} with respect to the small perturbations and performing the Fourier transform, the equation governing the fluctuation $\delta \tilde f(\omega,\mathbf{k},\mathbf{p})$ simplifies to
\begin{equation}
    \left[\left(\frac{1}{\tau_R}-i\omega+i\mathbf{k}\cdot\mathbf{v}\right)+B_0\left(v_z\frac{\partial}{\partial p_y}-v_y\frac{\partial}{\partial p_z}\right)\right]\delta \tilde f(\omega,\mathbf{k},\mathbf{p})=\frac{1}{\tau_R}\delta \tilde f_{\mathrm{eq}}(\omega,\mathbf{k},\mathbf{p})+\frac{f_0(\mathbf{p})}{T_0}\mathbf{v}\cdot \tilde{\mathbf{E}}(\omega,\mathbf{k}),   \label{RTA equation2}
\end{equation}
where we have explicitly written the Lorentz force operator  induced by the background magnetic field as $B_0\left(v_z\frac{\partial}{\partial p_y}-v_y\frac{\partial}{\partial p_z}\right)$ and defined the particle three-velocity as $\mathbf{v}=\frac{\mathbf{p}}{p^0}$.

%%%%%%%%%%%%%%%%%%%%%%%%%%%%%%%%%%%%%%%%%%%%%%
\section{Exact Solution of the Linearized Boltzmann Equation in a Uniform Magnetic Field}
\label{sec3 Exact Solution}
%%%%%%%%%%%%%%%%%%%%%%%%%%%%%%%%%%%%%%%%%%%%%%
The linearized relativistic Boltzmann equation \eqref{RTA equation2} represents a linear, inhomogeneous first-order partial differential equation in momentum space. To facilitate its analytical solution, it is convenient to introduce the following shorthand notations:
\begin{equation}
    D(\omega,\mathbf{k},\mathbf{p})=\frac{1}{\tau_R}-i\omega+i\mathbf{k}\cdot\mathbf{v},
\end{equation}
\begin{equation}
    \Delta(\omega,\mathbf{k},\mathbf{p})=\frac{1}{\tau_R}\delta \tilde f_{\mathrm{eq}}(\omega,\mathbf{k},\mathbf{p})+\frac{f_0(\mathbf{p})}{T_0}\mathbf{v}\cdot \tilde{\mathbf{E}}(\omega,\mathbf{k}).
\end{equation}
With these notations, Eq.\eqref{RTA equation2} can be written as
\begin{equation}
    \left[D(\omega,\mathbf{k},\mathbf{p})+B_0\left(v_z\frac{\partial}{\partial p_y}-v_y\frac{\partial}{\partial p_z}\right)\right]\delta \tilde f(\omega,\mathbf{k},\mathbf{p})=\Delta(\omega,\mathbf{k},\mathbf{p}).   \label{RTA equation3}
\end{equation}

To exploit the cylindrical symmetry induced by the external magnetic field, we solve the kinetic equation \eqref{RTA equation3} by transforming the transverse momentum components $(p_y, p_z)$ into polar coordinates
\begin{equation}
p_y = p_\perp \cos\phi, 
\qquad
p_z = p_\perp \sin\phi,
\end{equation}
where $p_\perp \equiv \sqrt{p_y^2 + p_z^2}$ denotes the transverse momentum. The corresponding particle velocity components are
\begin{equation}
v_x = \frac{p_x}{p^0},
\qquad
v_y = \frac{p_\perp}{p^0}\cos\phi,
\qquad
v_z = \frac{p_\perp}{p^0}\sin\phi.
\end{equation}
The magnetic-field-induced differential operator appearing in Eq.~\eqref{RTA equation3},
\begin{equation}
v_z \frac{\partial}{\partial p_y}
-
v_y \frac{\partial}{\partial p_z},
\end{equation}
can be rewritten in polar coordinates using the chain rule. The momentum derivatives are
\begin{align}
\frac{\partial}{\partial p_y}
&=
\cos\phi\,\frac{\partial}{\partial p_\perp}
-
\frac{\sin\phi}{p_\perp}\frac{\partial}{\partial\phi},
\\
\frac{\partial}{\partial p_z}
&=
\sin\phi\,\frac{\partial}{\partial p_\perp}
+
\frac{\cos\phi}{p_\perp}\frac{\partial}{\partial\phi}.
\end{align}
Substituting these expressions together with the velocity components, we obtain
\begin{align}
v_z \frac{\partial}{\partial p_y}
-
v_y \frac{\partial}{\partial p_z}
&=
\frac{p_\perp}{p^0}
\left[
\sin\phi
\left(
\cos\phi\,\partial_{p_\perp}
-
\frac{\sin\phi}{p_\perp}\partial_\phi
\right)
-
\cos\phi
\left(
\sin\phi\,\partial_{p_\perp}
+
\frac{\cos\phi}{p_\perp}\partial_\phi
\right)
\right]
\nonumber\\
&=
-\frac{1}{p^0}\frac{\partial}{\partial\phi}.
\end{align}
All radial derivatives cancel exactly, and the operator reduces to a derivative with respect to the polar angle $\phi$. This reduction reflects the fact that a uniform magnetic field changes only the direction of the transverse momentum, while leaving its magnitude unchanged.

Consequently, the linearized Boltzmann equation \eqref{RTA equation2} is simplified to
\begin{equation}
\left(
\frac{1}{\tau_R}
-
i\omega
+
i\mathbf{k}\cdot\mathbf{v}
-
\frac{B_0}{p^0}\frac{\partial}{\partial\phi}
\right)
\delta\tilde f(\omega,\mathbf{k},p_x,p_\perp,\phi)
=
\Delta(\omega,\mathbf{k},p_x,p_\perp,\phi).
\label{eq:polar_ode0}
\end{equation}
Rearranging terms yields a standard first-order linear ordinary differential equation in $\phi$,
\begin{equation} 
\frac{\partial}{\partial\phi}\delta\tilde f
-
\frac{p^0}{B_0}
D(\phi)\,
\delta\tilde f
=
-\frac{p^0}{B_0}\Delta(\phi),
\label{eq:polar_ode}
\end{equation}
where
\begin{equation}
D(\phi)
=
\frac{1}{\tau_R}
-
i\omega
+
i k_x v_x
+
i\frac{p_\perp}{p^0}
\left(
k_y\cos\phi
+
k_z\sin\phi
\right),
\end{equation}
and
\begin{equation}
\Delta(\phi)
=
\frac{f_0}{T_0}
\left[
\frac{\delta \tilde \mu}{\tau_R}
+
\frac{p_x}{p^0}\tilde E_x
+
\frac{p_\perp}{p^0}
\left(
\tilde E_y\cos\phi
+
\tilde E_z\sin\phi
\right)
\right].
\end{equation}

Eq.\eqref{eq:polar_ode} can be solved analytically by employing the method of integrating factors, whose general solution can be written as
\begin{equation}
\delta \tilde f(p_x,p_\perp,\phi)
=
\exp\!\left[
\frac{p_0}{B_0}
\int_{0}^{\phi} d\phi'\,D(\phi')
\right]
\left\{
C(p_x,p_\perp)
-
\frac{p_0}{B_0}
\int_{0}^{\phi}
d\phi'\,
\exp\!\left[
-\frac{p_0}{B_0}
\int_{0}^{\phi'} d\phi''\,D(\phi'')
\right]
\Delta(\phi')
\right\},
\label{eq:general_solution_phi}
\end{equation}
where the lower integration limit has been fixed to $\phi=0$ for convenience, and
$C(p_x,p_\perp)$ is an integration constant independent of the polar angle $\phi$.

Since the polar angle $\phi$ parametrizes a closed cyclotron orbit in the transverse
momentum plane $(p_y,p_z)$, the distribution function must be single-valued.
This requirement imposes the periodicity condition
\begin{equation}
\delta \tilde f(p_x,p_\perp,\phi+2\pi)
=
\delta \tilde f(p_x,p_\perp,\phi),
\label{eq:periodicity}
\end{equation}
which uniquely fixes the integration constant $C(p_x,p_\perp)$.

To implement this condition, we introduce the shorthand notation
\begin{equation}
\mathcal{I}(\phi)
\equiv
\frac{p_0}{B_0}
\int_{0}^{\phi} d\phi'\,D(\phi').
\label{I}
\end{equation}
Applying the boundary condition \eqref{eq:periodicity} to the solution \eqref{eq:general_solution_phi} yields the identity
\begin{align}
e^{\mathcal{I}(\phi+2\pi)}
\left[
C(p_x,p_\perp)
-
\frac{p_0}{B_0}
\int_{0}^{\phi+2\pi}
d\phi'\,
e^{-\mathcal{I}(\phi')}\Delta(\phi')
\right]=
e^{\mathcal{I}(\phi)}
\left[
C(p_x,p_\perp)
-
\frac{p_0}{B_0}
\int_{0}^{\phi}
d\phi'\,
e^{-\mathcal{I}(\phi')}\Delta(\phi')
\right].
\end{align}
Dividing both sides by $e^{\mathcal{I}(\phi)}$ and rearranging terms, one finds
\begin{equation}
\left(e^{\mathcal{I}(2\pi)}-1\right) C(p_x,p_\perp)
=
e^{\mathcal{I}(2\pi)}
\frac{p_0}{B_0}
\int_{0}^{2\pi}
d\phi\,
e^{-\mathcal{I}(\phi)}\Delta(\phi),
\end{equation}
where we have used the identity
$\mathcal{I}(\phi+2\pi)=\mathcal{I}(\phi)+\mathcal{I}(2\pi)$.
As long as $e^{\mathcal{I}(2\pi)}\neq 1$, which is ensured by the finite relaxation rate, the integration constant is uniquely determined as
\begin{equation}
C(p_x,p_\perp)
=
\frac{e^{\mathcal{I}(2\pi)}}{e^{\mathcal{I}(2\pi)}-1}
\;
\frac{p_0}{B_0}
\int_{0}^{2\pi}
d\phi\,
\exp\!\left[
-\frac{p_0}{B_0}
\int_{0}^{\phi} d\phi'\,D(\phi')
\right]
\Delta(\phi).
\label{eq:C_determined}
\end{equation}

Therefore, the requirement of single-valuedness completely removes the freedom associated with the integration constant. Substituting the explicit form of $D(\phi)$ into Eq.~\eqref{I}, one obtains
\begin{equation}
\mathcal{I}(\phi)=
\frac{1}{\omega_c}
\int_{0}^{\phi} d\phi'\,D(\phi')
= \frac{\alpha}{\omega_c}\phi + i \frac{p_\perp}{B_0} \left[ k_y \sin\phi + k_z (1 - \cos\phi) \right]= \frac{\alpha}{\omega_c}\phi + i \frac{p_\perp}{B_0}k_z+i\xi\sin(\phi-\psi),
\end{equation}
where $\psi \equiv \arctan(k_z/k_y)$ denotes the azimuthal angle of the transverse wave vector $\mathbf{k}_\perp=(k_y,k_z)$, with magnitude $k_\perp=\sqrt{k_y^2+k_z^2}$. We further introduce the complex frequency $\alpha \equiv \frac{1}{\tau_R}-i\omega+i k_x v_x$, the relativistic cyclotron frequency $\omega_c \equiv B_0/p^0$, and the dimensionless parameter $\xi\equiv k_\perp r_L$, where $r_L\equiv {p_\perp}/{B_0}$ is the relativistic Larmor radius characterizing the radius of the particle cyclotron orbit in the transverse plane under a uniform magnetic field. Using the Jacobi–Anger expansion
\begin{equation}
    e^{i z \sin\theta} = \sum_{n=-\infty}^{\infty} J_n(z) e^{i n \theta},
\end{equation}
where $J_n(z)$ is the Bessel function of the first kind, we obtain the exact, closed-form series solution of Eq.~\eqref{RTA equation2}
\begin{equation}
    \delta\tilde{f}(p_x,p_\perp, \phi) = \sum_{l=-\infty}^{\infty} \sum_{n=-\infty}^{\infty} \sum_{m=-1}^1 \frac{\Delta_m J_n(\xi) J_l(\xi)}{\frac{1}{\tau_R} - i(\omega - k_x v_x) + i(n-m)\omega_c} e^{i(l+m-n)\phi} e^{-i(l-n)\psi}, 
    \label{Exact Series Solution}
\end{equation}
where we have defined
\begin{align}
    \begin{aligned}
    \Delta(\phi) &= \sum_{m=-1}^{1} \Delta_m e^{im\phi},\\
    \Delta_0 &= \frac{f_0}{T_0} \left( \frac{\delta\tilde{\mu}}{\tau_R} + v_x \tilde{E}_x \right),\\
    \Delta_1 &= \frac{f_0}{T_0} \frac{p_\perp}{2p_0} (\tilde{E}_y - i\tilde{E}_z),\\
    \Delta_{-1} &= \frac{f_0}{T_0} \frac{p_\perp}{2p_0} (\tilde{E}_y + i\tilde{E}_z).
    \end{aligned}
\end{align}
For completeness, a detailed derivation of this series solution using the Jacobi-Anger expansion is provided in Appendix~\ref{app:derivation}. 
%%%%%%%%%%%%%%%%%%%%%%%%%%%%%%%%%%%%%%%%%%%%%%
\section{Analytic Expressions of the Current-current Correlators}
\label{sec4 Correlators}
%%%%%%%%%%%%%%%%%%%%%%%%%%%%%%%%%%%%%%%%%%%%%%
In this section, we employ the exact analytic solution \eqref{Exact Series Solution} to evaluate the retarded current-current correlators $G^{\mu\nu}_{JJ}(\omega,\mathbf{k})$. These correlation functions characterize the linear response of the  vector current  $ J^\mu$ to a small external $U(1)$ gauge perturbation $\delta A^\nu$ in the presence of the background magnetic field. The fluctuation of the charge density $\delta n(\omega,\mathbf{k})$ is obtained by integrating the distribution function perturbation over the momentum space
\begin{align}
\begin{aligned}
    \delta n(\omega, \mathbf{k})&=\int \frac{d^3\mathbf{p}}{(2\pi)^3} \delta \tilde{f}(\omega, \mathbf{k}, \mathbf{p})\\
    &=\int \frac{p_{\perp}dp_{\perp} dp_xd\phi}{(2\pi)^3}\sum_{l=-\infty}^{\infty} \sum_{n=-\infty}^{\infty} \sum_{m=-1}^1 \frac{\Delta_m J_n(\xi) J_l(\xi)}{\frac{1}{\tau_R} - i(\omega - k_x v_x) + i(n-m)\omega_c} e^{i(l+m-n)\phi} e^{-i(l-n)\psi}\\
    &=\sum\!\!\!\!\!\!\!\!\int \sum_{m=-1}^1\frac{\Delta_m J_{l+m}(\xi) J_l(\xi)}{\frac{1}{\tau_R} - i(\omega - k_x v_x) + il\omega_c}  e^{im\psi},
    \label{fluctuating number density of charge}
\end{aligned}
\end{align}
where we have introduced the shorthand notation $\sum\!\!\!\!\!\!\int=\int \frac{p_{\perp}dp_{\perp} dp_x}{(2\pi)^{2}}\sum_{l=-\infty}^{\infty}$ and used the selection rule
\begin{equation}
    \int_0^{2\pi}d\phi e^{i(l+m-n)\phi}=(2\pi)\delta_{l+m-n,0}.
\end{equation}
Physically, the perturbation of the local chemical potential $\delta\tilde{\mu}$ is determined by the matching condition, which ensures that the charge density coincides with its local equilibrium value as given by the local equilibrium distribution. Within the relaxation-time approximation, this condition is required to prevent the collision term from introducing an artificial source or sink of the conserved $U(1)$ charge. Restricting our analysis to the isolated charge transport sector by neglecting couplings to the energy-momentum sector ($\delta T=0$, $\delta u^\mu=0$), the system is fully closed by this matching condition, yielding the linear relation
\begin{equation}
    \delta n=\chi\delta\tilde\mu, \label{matching condition}
\end{equation}
where the static charge susceptibility $\chi$ is defined as
\begin{equation}
\chi= \frac{\partial n_0}{\partial \mu_0} = \frac{\partial}{\partial \mu_0} \int \frac{d^3 \textbf{p}}{(2\pi)^3} f_0\left(\mathbf{p}\right) = \frac{e^{\frac{\mu_0}{T_0}}}{2\pi^2} m^2 K_2(\frac{m}{T_0}),
\end{equation}
where $K_{n}(x)$ is the modified Bessel function of the second kind~\cite{gradshteyn2014table}. Substituting Eq.~\eqref{matching condition} into Eq.~\eqref{fluctuating number density of charge} yields
\begin{align}
    \begin{aligned}
    \label{fluctuating number density1}
        \delta n(\omega, \mathbf{k})=&\left(1-\sum\!\!\!\!\!\!\!\!\int \frac{ J_{l}(\xi) J_l(\xi)}{\frac{1}{\tau_R} - i(\omega - k_x v_x) + il\omega_c}\frac{f_0(\mathbf{p})}{T_0\tau_R\chi}\right)^{-1}\times\left(\sum\!\!\!\!\!\!\!\!\int \frac{J_{l}(\xi) J_l(\xi)}{\frac{1}{\tau_R} - i(\omega - k_x v_x) + il\omega_c}  \frac{f_0(\mathbf{p})}{T_0}v_x\tilde{E}_x \right.\\  &+\sum\!\!\!\!\!\!\!\!\int \frac{J_{l+1}(\xi) J_l(\xi)e^{i\psi}}{\frac{1}{\tau_R} - i(\omega - k_x v_x) + il\omega_c}\frac{f_0(\mathbf{p})}{T_0} \frac{p_\perp}{2p_0} (\tilde{E}_y - i\tilde{E}_z)\\
        &\left.+\sum\!\!\!\!\!\!\!\!\int \frac{J_{l-1}(\xi) J_l(\xi)e^{-i\psi}}{\frac{1}{\tau_R} - i(\omega - k_x v_x) + il\omega_c}\frac{f_0(\mathbf{p})}{T_0} \frac{p_\perp}{2p_0} (\tilde{E}_y + i\tilde{E}_z)\right).
    \end{aligned}
\end{align}
Having determined the charge-density fluctuation, we next obtain the spatial components of the current fluctuation by taking the corresponding momentum moments of the distribution-function perturbation:
\begin{equation}
    \delta J^i(\omega, \mathbf{k})=\int \frac{d^3\mathbf{p}}{(2\pi)^3p^0}p^i \delta \tilde{f}(\omega, \mathbf{k}, \mathbf{p}).
\end{equation}
The resulting spatial components are
\begin{align}
    \begin{aligned}
    \label{eq:deltaJ1}
        \delta J^1(\omega,\mathbf{k})=\sum\!\!\!\!\!\!\!\!\int\sum_{m=-1}^{1}\frac{\Delta_m J_{l+m}(\xi) J_l(\xi)}{\frac{1}{\tau_R} - i(\omega - k_x v_x) + il\omega_c}  e^{im\psi}v_x,
    \end{aligned}
\end{align}
\begin{align}
    \begin{aligned}
        \delta J^2(\omega,\mathbf{k})=\sum\!\!\!\!\!\!\!\!\int\sum_{m=-1}^{1}\sum_{s=\pm 1}\frac{\Delta_m J_{l+m+s}(\xi) J_l(\xi)}{\frac{1}{\tau_R} - i(\omega - k_x v_x) + i(l+s)\omega_c}  e^{i(m+s)\psi}\frac{p_\perp}{2p_0},
    \end{aligned}
\end{align}
\begin{align}
    \begin{aligned}
    \label{eq:deltaJ3}
        \delta J^3(\omega,\mathbf{k})=\sum\!\!\!\!\!\!\!\!\int\sum_{m=-1}^{1}\sum_{s=\pm 1}(-is)\frac{\Delta_m J_{l+m+s}(\xi) J_l(\xi)}{\frac{1}{\tau_R} - i(\omega - k_x v_x) + i(l+s)\omega_c}  e^{i(m+s)\psi}\frac{p_\perp}{2p_0}.
    \end{aligned}
\end{align}
The retarded current-current correlators are defined by
\begin{equation}
G^{\mu\nu}_{JJ}(\omega,\mathbf{k})\equiv \frac{\delta J^\mu}{\delta A_\nu}.
\end{equation}
Using the expressions for the charge-density and spatial-current fluctuations, Eqs.~\eqref{fluctuating number density1} and~\eqref{eq:deltaJ1}--\eqref{eq:deltaJ3}, we derive the following exact compact result for the retarded current-current correlators 
\begin{equation} \label{eq:compact_G}
G_{JJ}^{\mu\nu}(\omega,\mathbf{k})
=
-\chi\,\delta^\mu_0\delta^\nu_0
+
\omega
\left[
\mathcal{K}^{\mu\nu}
+
\frac{i}{\tau_R\chi}
\frac{
\mathcal{K}^{\mu0}\mathcal{K}^{0\nu}
}
{
1-\frac{i}{\tau_R\chi}\mathcal{K}^{00}
}
\right].
\end{equation}
Here, the kernel $\mathcal K^{\mu\nu}$ is given by
\begin{equation}
\mathcal{K}^{\mu\nu}(\omega,\mathbf{k})
\equiv
\sum_{l=-\infty}^{+\infty}
\int\frac{d^3\mathbf p}{(2\pi)^3}
\frac{f_0(\mathbf p)}{T_0}
\frac{
\mathcal{L}_l^\mu(\mathbf k,\mathbf p)
\mathcal{R}_l^\nu(\mathbf k,\mathbf p)
}
{
\omega-\Omega_l(\mathbf p)+i/\tau_R
},
\end{equation}
where the functions $\mathcal{L}_l^\mu$ and $\mathcal{R}_l^\nu$ are defined by
\begin{align}
\mathcal{L}_l^0
&\equiv
J_l(\xi),
\\
\mathcal{L}_l^1
&\equiv
v_xJ_l(\xi),
\\
\mathcal{L}_l^2
&\equiv
\frac{v_\perp}{2}
\left[
e^{i\psi}J_{l-1}(\xi)
+
e^{-i\psi}J_{l+1}(\xi)
\right],
\\
\mathcal{L}_l^3
&\equiv
\frac{i v_\perp}{2}
\left[
e^{-i\psi}J_{l+1}(\xi)
-
e^{i\psi}J_{l-1}(\xi)
\right],
\\
\mathcal{R}_l^0
&\equiv
J_l(\xi),
\\
\mathcal{R}_l^1
&\equiv
v_xJ_l(\xi),
\\
\mathcal{R}_l^2
&\equiv
\frac{v_\perp}{2}
\left[
e^{i\psi}J_{l+1}(\xi)
+
e^{-i\psi}J_{l-1}(\xi)
\right],
\\
\mathcal{R}_l^3
&\equiv
\frac{i v_\perp}{2}
\left[
e^{-i\psi}J_{l-1}(\xi)
-
e^{i\psi}J_{l+1}(\xi)
\right].
\end{align}
Here, we define $v_\perp \equiv p_\perp/p^0$ and $\Omega_l(\mathbf{p}) \equiv (k_xp_x+lB_0)/p^0$; the physical meaning of $\Omega_l(\mathbf{p})$ will be discussed later. In deriving and simplifying the individual components of the retarded correlator, we use the recurrence relations and the summation identity for the Bessel functions of the first kind given in Appendix~\ref{app:bessel_functions}, Eqs.~\eqref{eq:bessel_recurrence} and~\eqref{eq:bessel_summation}. For completeness, the explicit expressions for all sixteen components are collected in Appendix~\ref{app:explicit_correlators}.

It is straightforward to verify that the above analytic results identically fulfill the Ward identities:
\begin{equation}
    k_{\nu} G^{\mu\nu}_{JJ} = \omega G^{\mu0}_{JJ} - k_x G^{\mu1}_{JJ}- k_y G^{\mu2}_{JJ}- k_z G^{\mu3}_{JJ} = 0,
\end{equation}
\begin{equation}
    k_{\mu} G^{\mu\nu}_{JJ} = \omega G^{0\nu}_{JJ} - k_x G^{1\nu}_{JJ}- k_y G^{2\nu}_{JJ}- k_z G^{3\nu}_{JJ} = 0.
\end{equation}
These identities provide a consistency check for our analytical solutions. Physically, they reflect the local $U(1)$ gauge invariance and the macroscopic conservation of the vector current ($\partial_\mu \delta J^\mu = 0$). In linear response theory, they ensure that external gauge field perturbations do not violate the continuity equation, even in a background magnetic field. 
%%%%%%%%%%%%%%%%%%%%%%%%%%%%%%%%%%%%%%%%%%%%%%
\section{Analytic Structure of the Current-current Correlators}
\label{sec5 Analytic Structure}
%%%%%%%%%%%%%%%%%%%%%%%%%%%%%%%%%%%%%%%%%%%%%%

Although all components of $G_{JJ}^{\mu\nu}$ have been obtained above, we focus below on $G_{JJ}^{00}$ for the detailed analysis of the analytic structure. As the density-density retarded correlator, $G_{JJ}^{00}$ describes the response of the charge density to a perturbation of the temporal component of the gauge potential, and therefore provides a natural channel for probing the excitations in the charge-transport sector.  Moreover, $G_{JJ}^{00}$ already contains both the pole structure associated with the  hydrodynamic diffusion mode and the nonanalytic structure originating from the nonhydrodynamic sector,  as will be shown explicitly in the following analysis. For this reason, we restrict the detailed discussion below to $G_{JJ}^{00}$, while the other components can be analyzed analogously using the expressions given above.

\subsection{Hydrodynamic Limit}
In this subsection, we investigate the analytic structure of the retarded density-density correlator $G^{00}_{JJ}(\omega, \mathbf{k})$ in the hydrodynamic limit, i.e., the low-frequency and long-wavelength regime ($\omega \to 0$, $\mathbf{k} \to 0$). The collective modes of the system are determined by the poles of the retarded correlator in the complex $\omega$-plane. From the exact analytical expression \eqref{eq:compact_G}, the pole equation is obtained by setting the denominator of $G^{00}_{JJ}(\omega, \mathbf k)$ to zero, i.e.,
\begin{equation}
   1-\sum\!\!\!\!\!\!\!\!\int \frac{ J_{l}(\xi) J_l(\xi)}{\frac{1}{\tau_R} - i(\omega - k_x v_x) + il\omega_c}\frac{f_0(\mathbf{p})}{T_0\tau_R\chi}=0.
    \label{eq:pole_condition}
\end{equation}

To systematically extract the hydrodynamic dispersion relations, we employ a perturbative expansion in the low-frequency and long-wavelength regime. We introduce a formal dimensionless bookkeeping parameter $\epsilon \ll 1$ and scale the wave vector as $\mathbf{k} \to \epsilon \mathbf{k}$. We postulate a power-series ansatz for the dispersion relation 
\begin{equation}
    \omega(\epsilon) = \epsilon \omega_1 + \epsilon^2 \omega_2 + \mathcal{O}(\epsilon^3).
\end{equation}
Substituting this ansatz into Eq.~\eqref{eq:pole_condition}, and using the small-argument expansions of the Bessel functions given in Appendix~\ref{app:bessel_functions}, Eq.~\eqref{eq:bessel_small_argument}, we expand both the numerator and the denominator in powers of $\epsilon$:
\begin{align}
\begin{aligned}
&\frac{J_{0}(\epsilon\xi) J_0(\epsilon\xi)}{\frac{1}{\tau_R} - i(\epsilon \omega_1 + \epsilon^2 \omega_2+ \mathcal{O}(\epsilon^3)) + i \epsilon k_x v_x }\\&=\tau_R + i \tau_R^2 (-k_x v_x + \omega_1) \epsilon + \left[ -\frac{1}{2} \xi^2 \tau_R + i \tau_R^2 \left(i \tau_R \left(-k_x v_x + \omega_1\right)^2 + \omega_2\right) \right] \epsilon^2 + \mathcal{O}(\epsilon^3),
\end{aligned}
\end{align}
\begin{equation}
\frac{ J_{1}(\epsilon\xi) J_1(\epsilon\xi)}{\frac{1}{\tau_R} - i(\epsilon \omega_1 + \epsilon^2 \omega_2+ \mathcal{O}(\epsilon^3)) + i \epsilon k_x v_x + i\omega_c}=\frac{\xi^2 \tau_R}{4 + 4 i \tau_R \omega_c} \epsilon^2 + O(\epsilon^3),
\end{equation}
\begin{equation}
\frac{ J_{-1}(\epsilon\xi) J_{-1}(\epsilon\xi)}{\frac{1}{\tau_R} - i(\epsilon \omega_1 + \epsilon^2 \omega_2+ \mathcal{O}(\epsilon^3)) + i \epsilon k_x v_x - i\omega_c}=\frac{\xi^2 \tau_R}{4 - 4 i \tau_R \omega_c} \epsilon^2 + \mathcal O(\epsilon^3),
\end{equation}
where we have only considered the cases $l=0$ and $l=\pm1$; higher values of $l$ contribute corrections of higher order in $\epsilon$ and are therefore neglected here. Inserting these expansions back into the pole equation \eqref{eq:pole_condition} generates a polynomial identity in $\epsilon$:
\begin{align}
\begin{aligned}
    &\left[i \tau_R \left(-k_x\langle v_x\rangle+\omega_1\right) \right]\epsilon \\&+ \left[ i \tau_R \omega_2 - \tau_R^2 (\omega_1^2 - 2\omega_1 k_x \langle v_x \rangle + k_x^2 \langle v_x^2 \rangle) - \frac{ k_\perp^2 }{2}\langle\frac{ v_\perp^2 }{\omega_c^2}\rangle + \frac{ k_\perp^2 }{2} \langle \frac{  v_\perp^2 }{\omega_c^2(1 + \omega_c^2 \tau_R^2)}\rangle \right] \epsilon^2 + \mathcal{O}(\epsilon^3)=0.
\end{aligned}
\end{align}
For convenience, we have denoted the momentum-space average over the equilibrium distribution $f_{0}(\mathbf p)$ as 
\begin{equation}
    \langle A\rangle=\frac{1}{\chi T_0}\int\frac{d^3\mathbf p}{(2\pi)^3}f_{0}(\mathbf p)A.
\end{equation}
This average is properly normalized: from the definition of the susceptibility, $\chi T_0=\int d^3\mathbf p\,f_0/(2\pi)^3$, and therefore $\langle 1\rangle=1$. This normalization explains why the $\mathcal O(\epsilon^0)$ term in the expanded pole equation vanishes identically. The requirement that the equation holds for arbitrary $\epsilon$ dictates that the coefficient of each $\epsilon^n$ term must vanish identically,
\begin{equation}
    i \tau_R \left(-k_x\langle v_x\rangle+\omega_1\right)=0,
\end{equation}
\begin{equation}
    i \tau_R \omega_2 - \tau_R^2 (\omega_1^2 - 2\omega_1 k_x \langle v_x \rangle + k_x^2 \langle v_x^2 \rangle) - \frac{ k_\perp^2 }{2}\langle\frac{ v_\perp^2 }{\omega_c^2}\rangle + \frac{ k_\perp^2 }{2} \langle \frac{  v_\perp^2 }{\omega_c^2(1 + \omega_c^2 \tau_R^2)}\rangle =0.
\end{equation}
Because $f_0(\mathbf p)$ is even under $p_x\to-p_x$, while $v_x$ is odd, the normalized equilibrium average satisfies $\langle v_x\rangle=0$. Hence the $\mathcal{O}(\epsilon)$ equation gives $i\tau_R\omega_1=0$, which implies $\omega_1=0$. This mathematically confirms the absence of propagating modes in the charge diffusion channel. At $\mathcal{O}(\epsilon^2)$, substituting $\omega_1=0$ gives
\begin{equation}
    i \omega_2 \tau_R - \tau_R^2 k_x^2 \langle v_x^2 \rangle - \frac{\tau_R^2 k_\perp^2 }{2}\langle\frac{ v_\perp^2 }{1+\omega_c^2\tau_R^2}\rangle=0. \label{eq:omega2_eq}
\end{equation}
Solving Eq.~(\ref{eq:omega2_eq}) for the second-order coefficient $\omega_2$, we arrive at the dispersion relation for the diffusive mode:
\begin{equation}
    \omega(\mathbf{k}) = -i D_{L} k_x^2 - i D_{T} k_\perp^2+ \mathcal{O}(k^4) , \label{eq:dispersion_result}
\end{equation}
\begin{comment}
\begin{equation}
    \omega(\mathbf{k}) = -i (\tau_R \langle v_x^2 \rangle) k_x^2 - i \left( \frac{\tau_R}{2}\langle\frac{ v_\perp^2 }{1 + \omega_c^2 \tau_R^2}\rangle \right) k_\perp^2 + \mathcal{O}(k^4). \label{eq:dispersion_result}
\end{equation}
\end{comment}
where the longitudinal and transverse diffusion coefficients are defined respectively as
\begin{equation}
    \frac{D_L}{\tau_R} \equiv \langle v_x^2 \rangle= \frac{1}{3} + {\hat m} \frac{G_{1,3}^{2,1} \left( \frac{\hat{m}^2}{4} \left| \begin{array}{c} 1 \\ -\frac{1}{2}, \frac{1}{2}, 0 \end{array} \right. \right)}{12K_2 ({\hat m})} + {\hat m}\frac{\pi}{6K_2 ({\hat m})},
\end{equation}
\begin{equation}
    \frac{D_T}{\tau_R} \equiv \frac{1}{2}\langle\frac{ v_\perp^2 }{1 + \omega_c^2 \tau_R^2}\rangle. \label{DT}
\end{equation}
Here, $G^{2,1}_{1,3}$ denotes the Meijer G-function~\cite{gradshteyn2014table} and $\hat m \equiv m/T_0$ is the dimensionless mass. Our exact analytical evaluation of the longitudinal diffusion coefficient $D_L$ is in full agreement with the results reported in Ref.~\cite{Lin:2025ehr,Bajec:2025dqm}. Furthermore, in the massless limit ($\hat{m} \to 0$), the expression reduces to $D_L = \tau_R/3$, perfectly reproducing the findings of Refs.~\cite{Romatschke:2015gic,Bajec:2024jez,Hu:2023elg}.

From a physical perspective, the dispersion relation \eqref{eq:dispersion_result} explicitly demonstrates the anisotropic nature of charge transport in a magnetized plasma. The transport parallel to the magnetic field, characterized by $D_L$, remains completely unaffected by the background field, as the Lorentz force does no work and cannot modify the longitudinal component of the particle momentum. In contrast, the transverse diffusion ($D_T$) is suppressed by the background magnetic field. Microscopically, this transport behavior is governed by the dimensionless parameter $\hat{B}\equiv B_0\tau_R/T_0$, which characterizes the relative importance of cyclotron motion and collisional relaxation for typical thermal particles. In the strong-field regime, typical thermal particles complete multiple Larmor orbits between successive collisions. Consequently, their effective transverse displacement becomes constrained by the Larmor radius. This naturally leads to the suppression factor appearing in the integrand of Eq.~\eqref{DT}, explicitly breaking the transport isotropy.

\begin{figure}[h!] 
	\centering
	\includegraphics[width=0.65\textwidth]{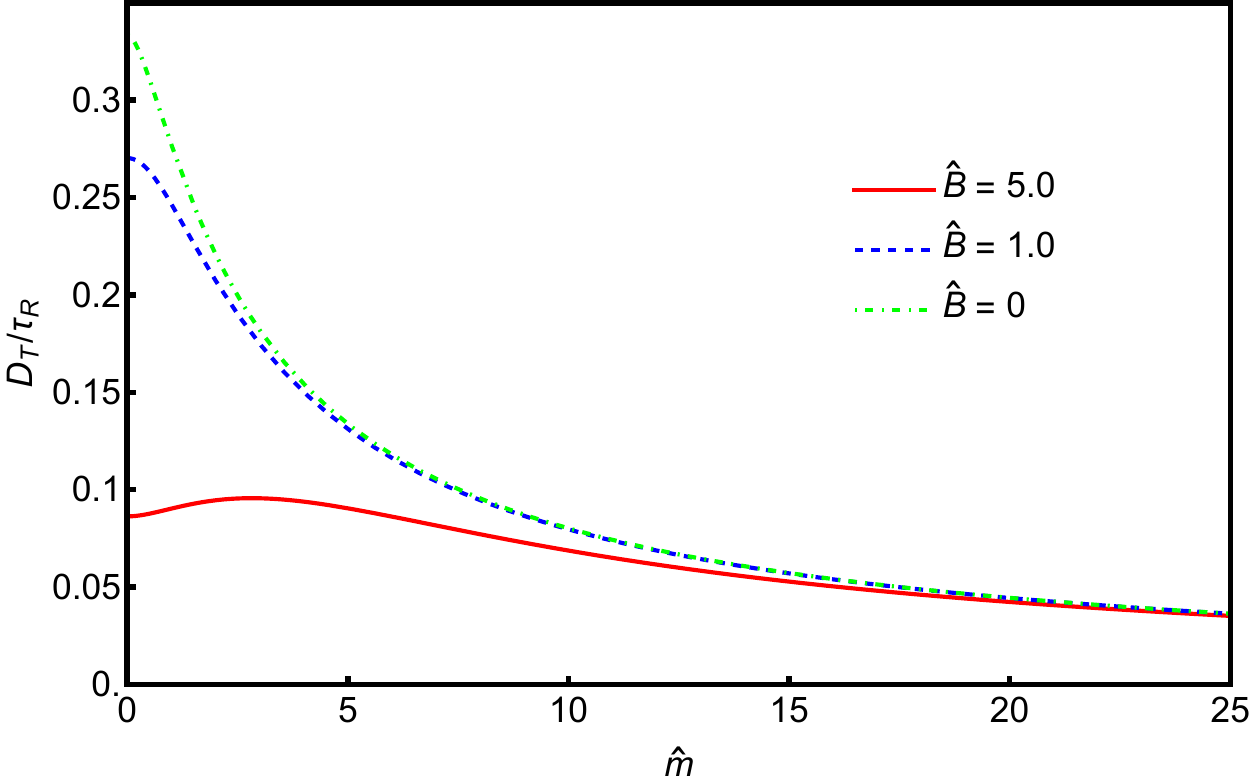}
	\caption{The dimensionless transverse diffusion coefficient $D_T/\tau_R$ as a function of the dimensionless mass $\hat{m} = m/T_0$ for different values of the background magnetic field ($\hat{B} = 0, 1.0, 5.0$).} 
	\label{fig1}
\end{figure}

\begin{figure}[h!] 
	\centering
	\includegraphics[width=0.65\textwidth]{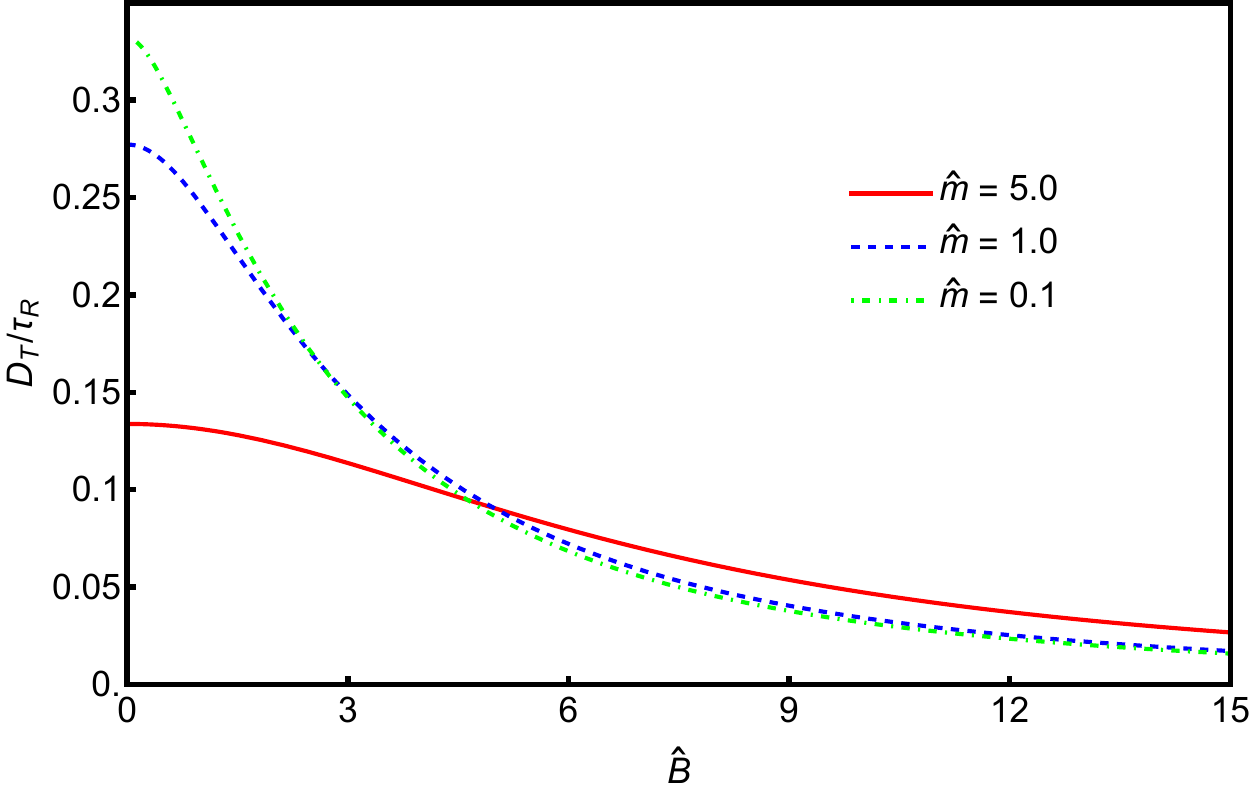}
	\caption{The dimensionless transverse diffusion coefficient $D_T/\tau_R$ as a function of the dimensionless magnetic field $\hat{B} = B_0 \tau_R / T_0$ for different mass parameters ($\hat{m} = 0.1, 1.0, 5.0$).} 
	\label{fig2}
\end{figure}

By introducing the dimensionless integration variable $x\equiv p/T_0$ together with the dimensionless magnetic-field parameter $\hat{B}$, Eq.~\eqref{DT} can be rewritten as
\begin{equation}
    \frac{D_T}{\tau_R} = \frac{1}{3 \hat{m}^2 K_2(\hat{m})} \int_{0}^{+\infty} dx\frac{x^4 e^{-\sqrt{x^2+\hat{m}^2}}}{x^2 + \hat{m}^2 + \hat{B}^2} . \label{DT_integral}
\end{equation}
The parametric dependence of the transverse diffusion coefficient $D_T$ on the particle mass and the magnetic-field strength is illustrated in Figs.~\ref{fig1} and \ref{fig2}. At $\hat{B}=0$, rotational symmetry gives $D_T=D_L$, and the decrease of this common diffusion coefficient with increasing $\hat{m}$ reflects the reduction of the thermal average of the squared particle speed. For nonzero $\hat{B}$, the mass dependence is additionally modified by the mass-dependent cyclotron-suppression factor $\left[1+(\omega_c\tau_R)^2\right]^{-1}$ and need not be monotonic. Figure~\ref{fig2} shows that, at fixed $\hat{m}$, increasing $\hat{B}$ suppresses the transverse diffusion. 
%%%%%%%%%%%%%%%%%%%%%%%%%%%%%%%%%%%%%%%%%%%%%%%%%%%%%
%%%%%%%%%%%%%%%%%%%%%%%%%%%%%%%%%%%%%%%%%%%%%%%%%%%%%

%%%%%%%%%%%%%%%%%%%%%%%%%%%%%%%%%%%%%%%%%%%%%%%%%%%%%
%%%%%%%%%%%%%%%%%%%%%%%%%%%%%%%%%%%%%%%%%%%%%%%%%%%%%

To further quantify this magnetic suppression, we extract analytical asymptotic behaviors from Eq.~\eqref{DT_integral}. At fixed $\hat{m}>0$, the weak-field expansion corresponds to the asymptotic limit $\hat{B}\to0$, which yields
\begin{equation}
\frac {D_T} {{\tau}_R}=\frac {D_L} {\tau_R} - \frac {\hat {B}^2} {2\hat {m}^2}\left[
    1 - \hat {m}\frac {K_ 1 (\hat {m})} {K_ 2 (\hat {m})} + (\hat{
{m}}^2 - 3)\frac {D_L} {\tau_R}  \right]+ \mathcal{O}\left({\hat{B}^4}\right).
\end{equation} Conversely, for fixed $\hat{m}$, the strong-field expansion is obtained in the asymptotic limit $\hat{B}\to\infty$. Since the expansion of the denominator in Eq.~\eqref{DT_integral} requires $\hat{B}^2\gg x^2+\hat{m}^2$ for the momenta that contribute substantially to the thermal integral, the relevant condition is $\hat{B}\gg\max\{1,\hat{m}\}$. Expanding the integrand in powers of $1/\hat{B}^2$ then gives
\begin{equation}
    \frac{D_T}{\tau_R} = \frac{\hat{m}}{\hat{B}^2} \frac{K_3(\hat{m})}{K_2(\hat{m})} - \frac{1}{\hat{B}^4} \left[ 5 \hat{m}^2 \frac{K_4(\hat{m})}{K_2(\hat{m})} + \hat{m}^3 \frac{K_3(\hat{m})}{K_2(\hat{m})} \right] + \mathcal{O}\left(\frac{1}{\hat{B}^6}\right).\label{DT_strong}
\end{equation}

\begin{figure}[h!] 
	\centering
	\includegraphics[width=0.65\textwidth]{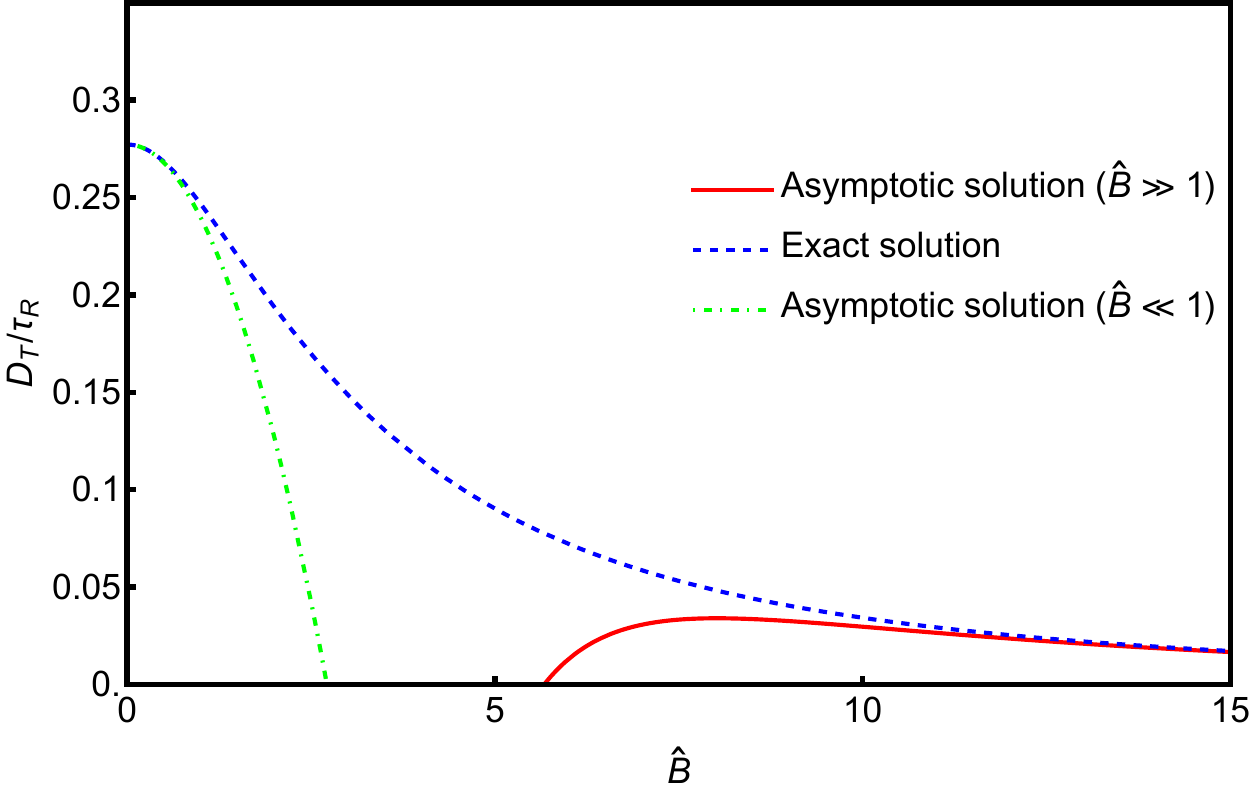}
	\caption{Comparison of the exact analytical calculation of the transverse diffusion coefficient (dashed blue line) with the weak-field asymptotic expansion (dashed green line) and the strong-field asymptotic expansion (solid red line). We choose the dimensionless mass $\hat{m}=1$. } %The expansions perfectly match the exact solution in their respective limits.
	\label{fig3}
\end{figure}

Physically, the analytical asymptotic behaviors derived above provide an explanation for the transport anisotropy and, more importantly, the reversal in the dependence of $D_T$ on the particle mass observed in Fig.~\ref{fig2}. In the weak-field limit, transport is dominated by the unhindered thermal motion of particles between collisions. Since lighter particles inherently possess a larger thermal average of the squared particle speed, they diffuse faster. Conversely, the leading-order $1/\hat{B}^2$ scaling in Eq.~\eqref{DT_strong} represents the transport suppression in strong magnetic-field environments. In this strong-field regime, the transverse transport can be viewed as a collision-driven two-dimensional random walk. The effective spatial step size of this random walk is constrained by the Larmor radius $r_L$, while the random shifts of the guiding center occur with an average time interval dictated by the relaxation time $\tau_R$. According to standard random walk theory, the macroscopic diffusion coefficient is determined by the squared step size per unit time. This picture yields $D_T \sim \langle r_L^2 \rangle / \tau_R = \langle p_\perp^2 \rangle / (B_0^2 \tau_R)$, simultaneously recovering the $1/\hat{B}^2$ suppression law and demonstrating that heavier particles—due to their larger mean squared thermal momentum $\langle p_\perp^2 \rangle$—exhibit a larger random-walk step size and consequently stronger transverse diffusion. The agreement between the exact integral and these asymptotic limits is illustrated in Fig.~\ref{fig3}.

%%%%%%%%%%%%%%%%%%%%%%%%%%%%%%%%%%%%%%%%%%%%%%%%%%%%%
%%%%%%%%%%%%%%%%%%%%%%%%%%%%%%%%%%%%%%%%%%%%%%%%%%%%%

\subsection{Weak-coupling Limit}
In this subsection, we investigate the weak-coupling (or rare-collision) regime, characterized by a parametrically large relaxation time $\tau_R T_0 \gg 1$. We perform a perturbative expansion in powers of the inverse relaxation time $1/\tau_R$,
\begin{equation} \label{perturbative expansion}
    G_{JJ}^{00}(\omega,\mathbf{k}) = G_0(\omega,\mathbf{k}) + G_1(\omega,\mathbf{k})/{\tau_R} + \mathcal{O}(1/\tau_R^2),
\end{equation}
where the leading-order term $G_0(\omega,\mathbf k)$ represents the purely collisionless Vlasov response,
\begin{equation}
    G_0(\omega, \mathbf{k}) = -\chi + \omega \int \frac{d^3\mathbf p}{(2\pi)^3} \frac{f_0(\mathbf p)}{T_0} \sum_{l=-\infty}^{+\infty} \frac{J_l(\xi)^2}{\omega - k_x v_x - l\omega_c + i0^+}.
    \label{eq:G00_LO}
\end{equation}
The first-order correction $G_1(\omega,\mathbf k)$, which captures the leading dissipative effects induced by rare scatterings, reads
\begin{align}
\begin{aligned}
    G_1(\omega,\mathbf{k}) =& {i\omega} \left\{ \frac{1}{\chi} \left[ \int \frac{d^3\mathbf p}{(2\pi)^3} \frac{f_0(\mathbf p)}{T_0} \sum_{l=-\infty}^{\infty} \frac{J_l(\xi)^2}{\omega - k_x v_x - l\omega_c+i0^+} \right]^2 \right.\\&\left.- \int \frac{d^3\mathbf p}{(2\pi)^3} \frac{f_0(\mathbf p)}{T_0} \sum_{l=-\infty}^{\infty} \frac{J_l(\xi)^2}{(\omega - k_x v_x - l\omega_c+i0^+)^2} \right\} \\
    =&i \left[ \frac{\partial G_0}{\partial \omega} + \frac{G_0}{\omega} \left( 1 + \frac{G_0}{\chi} \right) \right].
    \label{eq:G00_NLO_raw}
\end{aligned}
\end{align}

In linear response theory, the imaginary part of the retarded correlator encodes the spectral density and is related to the dissipation induced by the external perturbation. Therefore, to isolate the dissipative contributions to the charge transport, we extract the imaginary part of the retarded correlator in the weak-coupling limit below. We first define the fundamental complex integral appearing in the collisionless response~\eqref{eq:G00_LO} as 
\begin{equation}
    I_0(\omega, \mathbf{k}) \equiv \int \frac{d^3\mathbf{p}}{(2\pi)^3} \frac{f_0(\mathbf{p})}{T_0} \sum_{l=-\infty}^{+\infty} \frac{J_l(\xi)^2}{\omega - k_x v_x - l\omega_c + i0^+}.
\end{equation}
For real $\omega$, using the Sokhotski-Plemelj theorem, $1/(x+i0^+) = \mathcal{P}(1/x) - i\pi \delta(x)$, we can decompose this integral into its real part $I_{0,R}$ and its imaginary part $I_{0,I}$,
\begin{align}
    I_{0,R} &= \mathcal{P} \int \frac{d^3\mathbf{p}}{(2\pi)^3} \frac{f_0(\mathbf{p})}{T_0} \sum_{l=-\infty}^{+\infty} \frac{J_l(\xi)^2}{\omega - k_x v_x - l\omega_c}, \\
    I_{0,I} &= -\pi \int \frac{d^3\mathbf{p}}{(2\pi)^3} \frac{f_0(\mathbf{p})}{T_0} \sum_{l=-\infty}^{+\infty} J_l(\xi)^2 \delta(\omega - k_x v_x - l\omega_c).
\end{align}
Consequently, the leading-order correlator~\eqref{eq:G00_LO} can be written as $G_0(\omega, \mathbf{k}) = -\chi + \omega(I_{0,R} + iI_{0,I})$, which directly identifies the imaginary part as $\text{Im}\, G_0(\omega, \mathbf{k})/\omega = I_{0,I}$. 
Substituting this compact form into the first-order correction~\eqref{eq:G00_NLO_raw}, we obtain the imaginary part of the density-density correlator $G_{JJ}^{00}(\omega, \mathbf{k})$ up to $\mathcal{O}(1/\tau_R)$,
\begin{equation}
    \text{Im}\, G_{JJ}^{00}(\omega, \mathbf{k}) = \omega I_{0,I} + \frac{\omega}{\tau_R} \left[ \frac{\partial I_{0,R}}{\partial \omega} + \frac{1}{\chi} \left( I_{0,R}^2 - I_{0,I}^2 \right) \right] + \mathcal{O}(1/\tau_R^2).
\end{equation}

We now examine this response in the long-wavelength limit ($\mathbf{k} \to 0$). In this limit, we have
\begin{equation}
    \lim_{\mathbf{k} \to 0} \frac{\text{Im}\, G_{JJ}^{00}(\omega, \mathbf{k})}{\omega} = -\pi \chi \delta(\omega).
\end{equation}
This result can be understood through the Ward identity,
\begin{equation}
    k_\mu G_{JJ}^{\mu 0}(\omega, \mathbf{k}) = \omega G_{JJ}^{00}(\omega, \mathbf{k}) + k_i G_{JJ}^{i0}(\omega, \mathbf{k}) = 0.
\end{equation}
Since the current-density correlator $G_{JJ}^{i0}(\omega, \mathbf{k})$ remains finite as $\mathbf{k} \to 0$, it strictly follows that $\lim_{\mathbf{k} \to 0} \omega G_{JJ}^{00}(\omega, \mathbf{k}) = 0$. For this condition to hold for arbitrary frequencies, the imaginary part of $G_{JJ}^{00}(\omega, \mathbf{k})$ must strictly vanish for any finite frequency $\omega \neq 0$. Mathematically, this forces the ratio $\text{Im}\, G_{JJ}^{00} / \omega$ to be localized entirely at the origin, manifesting as a Dirac delta function $\delta(\omega)$. The coefficient $-\pi \chi$ is uniquely fixed by the Kramers-Kronig relations. Physically, the emergence of the Dirac delta function $\delta(\omega)$ in the long-wavelength limit ($\mathbf{k} \to 0$) is a manifestation of charge conservation. The  $\mathbf{k} \to 0$ limit corresponds to a spatially uniform perturbation of the charge density. Because the $U(1)$ charge is strictly conserved, such a uniform density fluctuation cannot decay or relax via local particle collisions, nor can it diffuse away, as there are no spatial gradients to drive a current. Consequently, this uniform mode has an infinite lifetime, meaning it cannot dissipate at any finite frequency. 

Furthermore, it is straightforward to verify that in the massless and zero-magnetic-field limits, the perturbative expansion \eqref{perturbative expansion} exactly reduces to
\begin{equation}
    G_{JJ}^{00}(\omega,k) = \chi \left( \frac{\omega}{2k} \ln \frac{\omega+k}{\omega-k} - 1 \right) + \frac{i \chi \omega}{\tau_R} \left[ \frac{1}{4k^2} \left( \ln \frac{\omega+k}{\omega-k} \right)^2 - \frac{1}{\omega^2 - k^2} \right] + \mathcal{O}(1/\tau_R^2),
\end{equation}
which recovers the analytical results previously derived in Ref.~\cite{Romatschke:2015gic}. We emphasize that the general weak-coupling expressions above do not require $m=0$; the massless limit is invoked here only as part of the simultaneous $m\to0$ and $B_0\to0$ consistency check.

%%%%%%%%%%%%%%%%%%%%%%%%%%%%%%%%%%%%%%%%%%%%%%
\subsection{Non-hydrodynamic Modes and Branch Cuts}
\label{sec:non_hydro_modes}
%%%%%%%%%%%%%%%%%%%%%%%%%%%%%%%%%%%%%%%%%%%%%%
In addition to the collective hydrodynamic excitations encoded by the poles of the retarded correlators, the kinetic theory in the relaxation time approximation encapsulates non-hydrodynamic modes. These modes physically represent the continuum of single-particle excitations and manifest mathematically as branch cuts in the complex $\omega$-plane~\cite{Hu:2024tnn,Lin:2025ehr,Bajec:2025dqm}. We restrict our attention to the correlator $G^{00}_{JJ}(\omega, \mathbf{k})$, and the other correlators can be analyzed in a similar way.

Specializing the retarded current-current correlators in Eq.~\eqref{eq:compact_G} to the density-density component and using $\mathcal{L}_l^0=\mathcal{R}_l^0=J_l(\xi)$, we introduce the notation
\begin{equation}
S(\omega,\mathbf{k})
\equiv \mathcal{K}^{00}(\omega,\mathbf{k})
=
\sum_{l=-\infty}^{\infty}S_l(\omega,\mathbf{k}),
\end{equation}
where
\begin{equation}
S_l(\omega,\mathbf{k})
=
\int\frac{d^3\mathbf{p}}{(2\pi)^3}
\frac{J_l^2(\xi)}
{\omega-\Omega_l(\mathbf{p})+\frac{i}{\tau_R}}
\frac{f_0(\mathbf{p})}{T_0}.
\label{eq:S_l_def}
\end{equation}
The retarded density-density correlator then takes the form
\begin{equation}
G_{JJ}^{00}(\omega,\mathbf{k})
=
-\chi
+
\frac{\omega S(\omega,\mathbf{k})}
{1-\frac{i}{\tau_R\chi}S(\omega,\mathbf{k})}.
\label{eq:G00_Mermin}
\end{equation}
The singular behavior of the integrand in Eq.~\eqref{eq:S_l_def} originates explicitly from its denominator $\omega - \Omega_l(\mathbf{p}) + \frac{i}{\tau_R}$. For a fixed physical momentum $\mathbf p$, the denominator vanishes at $\omega=\Omega_l(\mathbf p)-i/\tau_R$, giving a simple pole of the integrand whenever the numerator is nonzero. After integration over momentum, this continuum gives rise to branch cuts of the retarded correlator $G^{00}_{JJ}(\omega,\mathbf{k})$. The finite relaxation time shifts the cuts uniformly into the lower half-plane, where they lie along the horizontal line $\operatorname{Im}\omega=-1/\tau_R$ within the corresponding kinematically allowed ranges of $\operatorname{Re}\omega$. In what follows, we will further analyze the analytic structure and the discontinuity profile associated with these non-hydrodynamic cuts.

To extract the discontinuity across the cut, we introduce the following notation $\omega_\pm \equiv \operatorname{Re} \omega - i/\tau_R \pm i0^+$, where $\operatorname{Re} \omega \in \mathbb{R}$ denotes the real part of the cut. Substituting $\omega_\pm$ into Eq.~\eqref{eq:G00_Mermin}, the discontinuity profile is given by
\begin{align}
\begin{aligned}
    \text{disc} \, G_{JJ}^{00}(\omega,\mathbf{k})\big|_{\omega 
    =\operatorname{Re} \omega - i/\tau_R} &=G_{JJ}^{00}(\omega_+,\mathbf{k})-G_{JJ}^{00}(\omega_-,\mathbf{k}) \\
    &= \frac{\operatorname{Re} \omega - i/\tau_R}{\left( 1 - \frac{i}{\tau_R \chi} S(\omega_+,\mathbf{k}) \right) \left( 1 - \frac{i}{\tau_R \chi} S(\omega_-,\mathbf{k}) \right)} \sum_{l=-\infty}^{\infty} \Delta S_l(\operatorname{Re} \omega) \, ,
    \label{eq:disc_G00}
\end{aligned}
\end{align}
where $\Delta S_l(\operatorname{Re} \omega) \equiv S_l(\omega_+) - S_l(\omega_-)$ can be evaluated via the Sokhotski-Plemelj theorem as
\begin{align}
\begin{aligned}
    \Delta S_l(\operatorname{Re} \omega) &= -i \iint \limits_{p_x\in\mathbb{R},\; p_\perp \ge 0} dp_x dp_\perp \, h_l(p_x, p_\perp) \delta\big(\operatorname{Re} \omega - \Omega_l(p_x, p_\perp)\big) \, \\
    &= -i \int_{L_{\operatorname{Re} \omega}}  \frac{h_l(p_x, p_\perp)}{|\nabla \Omega_l(p_x, p_\perp)|} \, d\ell.
    \label{eq:Delta_Sl_2D}
\end{aligned}
\end{align}
Here, we have introduced the weight function
\begin{equation}
    h_l(p_x, p_\perp) \equiv \frac{p_\perp}{2\pi} J_l^2\left(\frac{k_\perp p_\perp}{B_0}\right) \frac{f_0(p_x, p_\perp)}{T_0} \, .
    \label{eq:weight_func}
\end{equation}
The Dirac delta function in Eq.~\eqref{eq:Delta_Sl_2D} restricts the two-dimensional integration to the contour
\begin{equation}
\label{Lre}
    L_{\operatorname{Re} \omega} = \left\{ (p_x, p_\perp) \in \mathbb{R} \times \mathbb{R}_{\ge 0} \;\big|\; \Omega_l(p_x, p_\perp) = \operatorname{Re} \omega \right\} ,
\end{equation}
where $d\ell$ is the infinitesimal line element on $L_{\operatorname{Re} \omega}$, and the gradient magnitude $|\nabla \Omega_l|$ emerges as the Jacobian factor.

Eq.~\eqref{eq:Delta_Sl_2D} offers a geometric perspective on the branch-cut structure. Since the equilibrium distribution function $f_0(\mathbf{p})$, the transverse momentum $p_\perp$, and the squared Bessel function are all non-negative, the weight function $h_l(p_x, p_\perp)$ is non-negative. The integral in Eq.~\eqref{eq:Delta_Sl_2D} can be nonzero only if the resonance condition $\operatorname{Re}\omega=\Omega_l(p_x,p_\perp)$ has solutions in the physical momentum domain with nonvanishing weight $h_l$. Thus, determining the location of the branch cuts is closely related to finding the global range of the mapping $\Omega_l: \mathbb{R} \times \mathbb{R}_{\ge 0} \to \mathbb{R}$.

For definiteness, we take $B_0>0$ and first consider the generic case $k_\perp\neq0$ and $k_x\neq0$. For $l\neq0$, a straightforward analysis of the stationary points of $\Omega_l(\mathbf p)$ yields the corresponding extremal value of $\Omega_l(\mathbf p)$:
\begin{equation}
    \Omega_l^{\text{ext}} = \text{sgn}(l) \sqrt{k_x^2 + l^2 \omega_{c0}^2} \, ,
    \label{eq:kinematic_threshold}
\end{equation}
where $\omega_{c0} \equiv B_0/m$ is the non-relativistic cyclotron frequency. Combining this extremal value with the ultra-relativistic asymptotic behavior $\lim_{|p_x| \to \infty} \Omega_l(\mathbf{p}) = \operatorname{sgn}(p_x) k_x$, we obtain the branch-cut interval associated with each harmonic channel:
\begin{itemize}
    \item 
    $l=0$: The branch cut contributed by this channel extends over the symmetric interval $\operatorname{Re} \omega \in (-\left|k_x\right|, \left|k_x\right|)$.
    \item 
    $l>0$: Bounded by the absolute maximum $\Omega_l^{\text{ext}}$, the branch cut for positive harmonics spans the interval $\operatorname{Re} \omega \in (-\left|k_x\right|, \sqrt{k_x^2 + l^2 \omega_{c0}^2} \, ]$.
    \item 
    $l<0$: Dictated by the symmetry, the branch cut for negative harmonics is precisely mirrored, occupying the range $\operatorname{Re} \omega \in [ \, -\sqrt{k_x^2 + l^2 \omega_{c0}^2}, \left|k_x\right|)$.
\end{itemize}
Because the harmonic index $l$ runs over all integers, the kinematic thresholds $\pm \sqrt{k_x^2 + l^2 \omega_{c0}^2}$ extend to arbitrarily large absolute values. Crucially, because the weight functions $h_l(p_x, p_\perp)$ for each harmonic channel are non-negative (see e.g., Eq.\eqref{eq:weight_func}),  their contributions to the total discontinuity add constructively and are not subject to nontrivial cancellations. Consequently, for the generic case $k_x\neq0$ and $k_\perp\neq0$, the
total branch cut of the correlator $G_{JJ}^{00}(\omega,\mathbf{k})$ lies
along the horizontal line
\begin{equation}
\operatorname{Re}\omega\in(-\infty,\infty),
\qquad
\operatorname{Im}\omega=-\frac{1}{\tau_R}.
\end{equation}

The special cases in which $k_\perp=0$ or $k_x=0$ require separate
consideration:
\begin{itemize}
    \item If $k_\perp=0$, then $\xi=0$ and
    $J_l(0)=\delta_{l,0}$. Hence only the $l=0$ channel remains. For $k_x\neq0$,
    the branch cut lies along
    \begin{equation}
    \operatorname{Re}\omega\in(-|k_x|,|k_x|),
    \qquad
    \operatorname{Im}\omega=-\frac{1}{\tau_R}.
    \end{equation}
    If, in addition, $k_x=0$, then in the strictly homogeneous limit one finds $G_{JJ}^{00}(\omega,\mathbf 0)=0$, reflecting charge conservation. No branch cut is generated for the retarded density-density correlator in this limit.
    \item If $k_x=0$ while $k_\perp\neq0$, then
$\Omega_0(\mathbf p)=0$ is independent of momentum. Thus, the $l=0$ channel does not generate a Landau-damping branch cut. For $l\neq0$, we have $\Omega_l(\mathbf p)=l\omega_c(\mathbf p)=\frac{lB_0}{p^0}$. The $l>0$ and $l<0$ channels contribute branch cuts over the intervals $\operatorname{Re}\omega\in(0,l\omega_{c0}]$ and $\operatorname{Re}\omega\in[-|l|\omega_{c0},0)$, respectively, all at $\operatorname{Im}\omega=-1/\tau_R$. As $|l|$ is unbounded, these cuts extend to arbitrarily large $|\operatorname{Re}\omega|$ and accumulate at the point $\omega=-i/\tau_R$. 
\end{itemize}

We further simplify $\Delta S_l$. Without loss of generality, we assume $B_0 > 0$ and $k_x > 0$ to facilitate the discussion. The derivation below initially assumes $\operatorname{Re}\omega\neq0$; the case $\operatorname{Re}\omega=0$ is treated separately later. Recalling that $\Omega_l(\mathbf p) = (k_x p_x + l B_0)/p^0$, the resonance condition forces the particle energy to be
\begin{equation}
    p^0 = \frac{k_x p_x + l B_0}{\operatorname{Re}\omega}.
\end{equation}
Since the physical particle energy must be strictly positive ($p^0 > 0$), this condition introduces the first Heaviside step function $\theta\left(\frac{k_x p_x + l B_0}{\operatorname{Re}\omega}\right)$. Furthermore, using the mass-shell condition $(p^0)^2 = p_x^2 + p_\perp^2 + m^2$, we can solve for the squared transverse momentum:
\begin{equation}
    p_\perp^2 = \left(\frac{k_x p_x + l B_0}{\operatorname{Re}\omega}\right)^2 - p_x^2 - m^2.
\end{equation}
The requirement that the transverse momentum $p_\perp$ must be a real number implies $p_\perp^2 \ge 0$, which explicitly yields the second Heaviside step function $\theta\left(\left(\frac{k_x p_x + l B_0}{\operatorname{Re}\omega}\right)^2-p_x^2-m^2\right)$. By integrating out the transverse momentum $p_\perp$ via the Dirac delta function $\delta\big(\operatorname{Re} \omega - \Omega_l(p_x, p_\perp)\big)$, Eq.~\eqref{eq:Delta_Sl_2D} is reduced to a one-dimensional integral over the longitudinal momentum $p_x$,
\begin{align}
\begin{aligned}
    &\Delta S_l(\operatorname{Re} \omega) \\&= -i \int_{-\infty}^{+\infty} dp_x \mathcal{H}_l(p_x,p_\perp)\, \theta\left(\frac{k_x p_x + l B_0}{\operatorname{Re}\omega}\right)\theta\left(\left(\frac{k_x p_x + l B_0}{\operatorname{Re}\omega}\right)^2-p_x^2-m^2\right)\Bigg|_{p_{\perp}=\sqrt{\left(\frac{k_x p_x + l B_0}{\operatorname{Re}\omega}\right)^2-p_x^2-m^2}},
\end{aligned}
\end{align}
where we have defined the function
\begin{align}
    \begin{aligned}
        \mathcal{H}_l(p_x,p_\perp)\equiv\frac{p_0^2h_l(p_x,p_\perp)}{\left|\operatorname{Re}\omega\right|p_\perp}=\frac{p_0^2}{2\pi T_0\left|\operatorname{Re}\omega\right|} J_l^2\left(\xi\right)f_0\left(\mathbf{p}\right).
    \end{aligned}
\end{align}
To explicitly determine the integration domain for $p_x$, we complete the square for the quadratic inequality imposed by the second Heaviside step function ($p_\perp^2 \ge 0$), rewriting it as
\begin{equation}
    \left[ (\operatorname{Re}\omega)^2 - k_x^2 \right] \left[ R^2 - (p_x - p_c)^2 \right] \ge 0.
    \label{second Heaviside step}
\end{equation}
Here, we naturally identify the frequency-dependent resonance center $p_c$ and the geometric momentum radius $R$, defined as
\begin{align}
    p_c &= \frac{lB_0k_x}{\left(\operatorname{Re}\omega\right)^2-k_x^2}, \\
    R &= \frac{\left|\operatorname{Re}\omega\right|m\sqrt{k_x^2+l^2\omega_{c0}^2-\left(\operatorname{Re}\omega\right)^2}}{\left|\left(\operatorname{Re}\omega\right)^2-k_x^2\right|}.
\end{align}
Eq.~\eqref{second Heaviside step} dictates both the kinematic bounds and the corresponding variable substitutions utilized below. We first consider $l>0$; the case $l=0$ is treated separately, while the case $l<0$ follows from the symmetry relation given below. For $0<|\operatorname{Re}\omega|<k_x$, the inequality becomes $(p_x-p_c)^2\geq R^2$, which gives two unbounded intervals. The first Heaviside function then selects the physically allowed momentum branch; this branch is naturally parameterized by the hyperbolic substitution $p_x=p_c\pm R\cosh\eta$ with $\eta\in[0,+\infty)$. Conversely, in the region $k_x<|\operatorname{Re}\omega|<\sqrt{k_x^2+l^2\omega_{c0}^2}$, the constraint $(p_x-p_c)^2\leq R^2$ forms a bounded interval, prompting the trigonometric substitution $p_x=p_c+R\sin\theta$ with $\theta\in[-\pi/2,\pi/2]$. 

At $\operatorname{Re}\omega=0$, the resonance condition reduces to $k_xp_x+lB_0=0$. Therefore Eq.~\eqref{eq:Delta_Sl_2D} gives
\begin{equation}
\Delta S_l(0)
=
-i\int_{0}^{+\infty}dp_\perp\,\frac{p_\perp}{2\pi k_x}\sqrt{p_\perp^2+m^2+\left(lB_0/k_x\right)^2}
\,
J^2_l(\xi)\frac{f_0(-\frac{lB_0}{k_x},p_\perp)}{T_0}.
\end{equation}

At $\operatorname{Re}\omega=k_x$, the parameters $p_c$ and $R$ become singular. Direct evaluation of Eq.~\eqref{eq:Delta_Sl_2D} yields
\begin{equation}
\begin{aligned}
\Delta S_l(k_x)
={}&-i
\int_{p_{x,\text{min}}^{(l)}}^{+\infty}
dp_x\,
\frac{\left(p_x+lB_0/k_x\right)^2}{2\pi k_x T_0}
J_l^2\left[
\frac{k_\perp}{B_0}
p_\perp^{(l)}(p_x)
\right]
f_0\left(
p_x,
p_\perp^{(l)}(p_x)\right),
\end{aligned}
\end{equation}
where the lower integration limit is defined as $p_{x,\min}^{(l)}=\dfrac{k_x}{2lB_0}\left(m^2-\dfrac{l^2B_0^2}{k_x^2}\right)$ and $p_\perp^{(l)}(p_x)=\sqrt{\dfrac{2lB_0}{k_x}p_x+\dfrac{l^2B_0^2}{k_x^2}-m^2}$.

For $l>0$, combining the hyperbolic and trigonometric substitutions with the direct evaluations at the  points $\operatorname{Re}\omega=0$ and $\operatorname{Re}\omega=k_x$, we obtain the following complete piecewise representation of $\Delta S_l$:
\begin{align}
\begin{aligned}
&\Delta S_l(\operatorname{Re}\omega) =\\
&\begin{cases}
-i \int_{0}^{+\infty} d\eta \, (R \sinh \eta) \, \mathcal{H}_l\left(p_x=p_c + R \cosh \eta, p_\perp=\frac{\sqrt{k_x^2-\left(\operatorname{Re}\omega\right)^2}}{\left|\operatorname{Re}\omega\right|} R \sinh \eta\right), & 0 < \operatorname{Re}\omega < k_x \\
-i \int_{0}^{+\infty} d\eta \, (R \sinh \eta) \, \mathcal{H}_l\left(p_x=p_c - R \cosh \eta, p_\perp=\frac{\sqrt{k_x^2-\left(\operatorname{Re}\omega\right)^2}}{\left|\operatorname{Re}\omega\right|} R \sinh \eta\right), & -k_x < \operatorname{Re}\omega < 0 \\
-i \int_{-\pi/2}^{\pi/2} d\theta \, (R \cos \theta) \, \mathcal{H}_l\left(p_x=p_c + R \sin \theta,  p_\perp=\frac{\sqrt{\left(\operatorname{Re}\omega\right)^2-k_x^2}}{\left|\operatorname{Re}\omega\right|} R \cos \theta\right), & k_x < \operatorname{Re}\omega < \sqrt{k_x^2+l^2\omega_{c0}^2} \\
-i\int_{0}^{+\infty}dp_\perp\,\frac{p_\perp}{2\pi k_x}\sqrt{p_\perp^2+m^2+\left(lB_0/k_x\right)^2}
\,
J^2_l(\xi)\frac{f_0(-\frac{lB_0}{k_x},p_\perp)}{T_0},  & \operatorname{Re}\omega=0\\
-i
\int_{p_{x,\text{min}}^{(l)}}^{+\infty}
dp_x\,
\frac{\left(p_x+lB_0/k_x\right)^2}{2\pi k_x T_0}
J_l^2\left[
\frac{k_\perp}{B_0}
p_\perp^{(l)}(p_x)
\right]
f_0\left(
p_x,
p_\perp^{(l)}(p_x)
\right)
, & \operatorname{Re}\omega=k_x\\
0, & \text{otherwise}
\end{cases}
\end{aligned}
\end{align}
For the case $ l = 0 $, the magnetic field entirely decouples from the resonance condition, leading to $p_c = 0$ and $R=\frac{\left|\operatorname{Re}\omega\right|m}{\sqrt{k_x^2-\left(\operatorname{Re}\omega\right)^2}}$. In this case, we have
\begin{align}
\begin{aligned}
\Delta S_0(\operatorname{Re}\omega) =
\begin{cases}
-i \int_0^{+\infty} d\eta \, (R \sinh \eta) \mathcal{H}_{0}\left(p_x=R \cosh \eta, p_\perp=m\sinh \eta\right), & 0 < \operatorname{Re}\omega < k_x \\
-i \int_0^{+\infty} d\eta \, (R \sinh \eta) \mathcal{H}_{0}\left(p_x=-R \cosh \eta, p_\perp=m\sinh \eta\right), & -k_x < \operatorname{Re}\omega < 0 \\
-i\int_{0}^{+\infty}dp_\perp\,\frac{p_\perp}{2\pi k_x}\sqrt{p_\perp^2+m^2}
\,
J^2_0(\xi)\frac{f_0(0,p_\perp)}{T_0},  & \operatorname{Re}\omega=0\\
0, & \text{otherwise}
\end{cases}
\end{aligned}
\end{align}
Finally, for the case $l < 0$, by exploiting the symmetry of the system, we can directly obtain
\begin{equation}
    \Delta S_{-l}\left(\operatorname{Re}\omega\right) = \Delta S_{l}\left(-\operatorname{Re}\omega\right).
\end{equation}

\subsection{Physical Interpretation of Branch Cuts and Cyclotron Damping}

To uncover the microscopic physical mechanisms underlying the non-hydrodynamic branch cuts, it is instructive to trace their mathematical emergence back to the singularities of the distribution function \eqref{Exact Series Solution}. As explicitly shown in the derivation of the discontinuity profile $\Delta S_l$ [e.g., Eq.~\eqref{eq:Delta_Sl_2D}], the non-analytic behavior of the retarded correlator $G^{00}_{JJ}(\omega,\mathbf{k})$ fundamentally stems from the singular poles of the momentum integral. After utilizing the Sokhotski--Plemelj theorem, these poles manifest as the Dirac delta function $\delta(\omega_r - k_x v_x - l\omega_c)$, where $\omega_r \equiv \text{Re } \omega$ denotes the real wave frequency.  This immediately dictates the relativistic wave-particle resonance condition,
\begin{equation}
    \omega_r - k_x v_x = l \omega_c, \quad l \in \mathbb{Z}. \label{resonance condition2}
\end{equation}

Physically, this condition reflects the interaction between charged particles in a background magnetic field $\mathbf{B}_0$ and an external plane-wave perturbation $\delta A_\mu$ with frequency $\omega_r$ and wave vector $\mathbf{k}$. When the resonance condition~\eqref{resonance condition2} is satisfied, the particles undergo a sustained energy exchange with the wave, which manifests as damping of the wave. Depending on the harmonic index $l$, these damping mechanisms can be classified into Landau damping and cyclotron damping:

The case of $l=0$ yields the resonance condition $\omega_r = k_x v_x$, which encapsulates Landau damping. This corresponds to a charged particle being continuously accelerated by the longitudinal wave electric field when its velocity component along the background magnetic field $\mathbf{B}_0$ matches the longitudinal phase velocity of the wave. As the particle moves in a circular orbit in the transverse plane, any energy it extracts from the transverse electric field of the wave during one phase of the cycle is transferred back to the field during the opposite phase. This cancellation prohibits any systematic, secular energy exchange in the transverse direction. Consequently, sustained energy transfer is strictly restricted to the longitudinal axis. Since the particle speed is physically bounded by the speed of light ($|v_x| < 1$), the branch cut contributed by this $l=0$ channel is strictly confined to the finite interval $\operatorname{Re} \omega \in (-|k_x|, |k_x|)$, as derived in Sec.~\ref{sec:non_hydro_modes}.

The channels with $l \neq 0$ represent cyclotron damping. The physical origin of these resonances, and the validity of Eq.~\eqref{resonance condition2}, can be understood by transforming to a longitudinally co-moving frame that translates along the background magnetic field $\mathbf{B}_0 = B_0 \hat{\mathbf{x}}$ with the particle's longitudinal velocity $v_x$. In this longitudinally co-moving frame, the particle's longitudinal motion is eliminated, and its trajectory reduces to a pure circular orbit in the transverse plane. The particle electrodynamically acts as an oscillator radiating at frequency $\Omega' = \gamma_x \omega_c$, where $\gamma_x = (1 - v_x^2)^{-1/2}$. Concurrently, the external gauge field perturbation with laboratory frequency $\omega_r$ and longitudinal wave number $k_x$ undergoes a relativistic Doppler shift. The wave frequency perceived by the particle in its co-moving frame is modified to $\omega'_r = \gamma_x (\omega_r - k_x v_x)$. Resonant energy exchange and subsequent kinetic damping occur when the Doppler-shifted wave frequency matches an integer multiple of the oscillator's frequency, namely $\omega'_r = l \Omega'$. Upon cancelling the common Lorentz factor $\gamma_x$, this resonance condition precisely yields the expression in Eq.~\eqref{resonance condition2}.

For a given harmonic index $l$, Eq.~\eqref{resonance condition2} determines a discrete resonant frequency for an individual particle. However, in a magnetized relativistic thermal medium described by the equilibrium distribution $f_0(\mathbf{p})$, the momentum distribution of charged particles is continuous. Consequently, for frequencies within the kinematically allowed range, a continuous set of resonant particles exists. The momentum integration over this thermal bath effectively connects and smears these infinitely dense, discrete poles into a continuous branch cut in the complex $\omega$-plane. The inclusion of the RTA collision term then uniformly shifts this entire continuous spectrum into the lower-half complex plane to $\operatorname{Im} \omega = -1/\tau_R$, reflecting the finite lifetime of these non-hydrodynamic excitations.

%%%%%%%%%%%%%%%%%%%%%%%%%%%%%%%%%%%%%%%%%%%%%%
\section{Summary and Outlook}
\label{sec6 Summary and Conclusion}
%%%%%%%%%%%%%%%%%%%%%%%%%%%%%%%%%%%%%%%%%%%%%%
In this work, we have investigated the charge transport properties of a magnetized relativistic plasma within the framework of kinetic theory. By employing the relaxation-time approximation, we solved the linearized relativistic Boltzmann equation in the presence of a constant and uniform background magnetic field. The cylindrical symmetry induced by the magnetic field allowed us to derive an analytic solution for the perturbation of the single-particle distribution function, which takes the form of a series expansion involving Bessel functions. Utilizing this solution, we computed the full set of retarded current-current correlators $G^{\mu\nu}_{JJ}(\omega,\mathbf{k})$ and verified that they satisfy the Ward identities, ensuring charge conservation and gauge invariance.

The analytic structure of these results enabled an exploration of various physical limits. In the hydrodynamic regime ($\omega \to 0$, $\mathbf{k} \to 0$), we extracted the dispersion relation of the collective diffusive mode. The results demonstrate the anisotropic nature of charge transport in magnetized environments: the longitudinal diffusion coefficient $D_L$ is unaffected by the magnetic field, whereas the transverse diffusion coefficient $D_T$ experiences a significant suppression. By analyzing the asymptotic behaviors, we showed that $D_T$ scales as $1/B_0^2$ in the strong-field limit. This scaling behavior can be understood through the microscopic picture of a collision-driven two-dimensional random walk constrained by the Larmor radius. Furthermore, we revealed that magnetic suppression induces a reversal in the mass dependence of transverse transport: the diffusion process transitions from an unhindered thermal motion regime, which facilitates the transport of lighter particles in weak fields, to a Larmor-radius-constrained random walk in which the transverse diffusion coefficient becomes larger for heavier particles at sufficiently strong fields.

Beyond the hydrodynamic limit, we analyzed the non-hydrodynamic modes of the system, which manifest as branch cuts in the complex frequency plane. We determined the kinematic thresholds of these continuous spectra and provided a physical interpretation based on relativistic wave-particle resonances. Specifically, the $l=0$ harmonic channel corresponds to Landau damping, while the $l \neq 0$ channels capture the cyclotron damping. 

Looking forward, several theoretical extensions can be considered based on the present framework. One direction is to generalize the constant relaxation-time approximation by introducing a momentum-dependent relaxation time; in that case, the non-analytic structure of the retarded correlators is expected to become more intricate. Another extension involves considering other types of external fields, such as systems with nonzero vorticity or in accelerating configurations that are also consistent with the Killing condition for global equilibrium, and  computing the associated retarded correlators under such fields. Finally, the spin tensor can be obtained by varying the action with respect to the corresponding source field. Thus, once a suitable kinetic equation incorporating spin degrees of freedom is formulated, the method and framework developed in this work can be applied to study the linear response behavior related to the spin tensor \cite{Hu:2022lpi,Hu:2022xjn,Hu:2022mvl}, which remains a subject for future investigation.

\acknowledgments

 We thank Pengfei Zhuang, Koichi Hattori, Xu-Guang Huang, Shuzhe Shi and Puyuan Bai for helpful discussions. This work was supported by Yantai University under Grant No.~2226001 and by the National Natural Science Foundation of China under Grant No.~12505149.

%%%%%%%%%%%%%%%%%%%%%%%%%%%%%%%%%%%%%%%%%%%%%%
\appendix
%%%%%%%%%%%%%%%%%%%%%%%%%%%%%%%%%%%%%%%%%%%%%%

%%%%%%%%%%%%%%%%%%%%%%%%%%%%%%%%%%%%%%%%%%%%%%
\section{Properties of Bessel Functions}
\label{app:bessel_functions}
%%%%%%%%%%%%%%%%%%%%%%%%%%%%%%%%%%%%%%%%%%%%%%
This appendix summarizes the definitions, recurrence relations, and asymptotic behaviors of the Bessel functions utilized in the main text.

\subsection{Bessel Functions of the First Kind}
\label{app:bessel_first}

The Bessel function of the first kind of integer order $n$, denoted as $J_n(z)$, is the canonical solution to Bessel's differential equation, $z^2 y'' + z y' + (z^2 - n^2)y = 0$. It can be explicitly defined by its Taylor series expansion around $z = 0$:
\begin{equation}
    J_n(z) = \sum_{k=0}^{\infty} \frac{(-1)^k}{k!\Gamma(k+n+1)} \left(\frac{z}{2}\right)^{2k+n}, \qquad n\in\mathbb{Z}_{\geq 0}.
\end{equation}
Negative integer orders are obtained from $J_{-n}(z)=(-1)^nJ_n(z)$. A fundamental identity connecting these functions to complex exponentials is the Jacobi--Anger expansion, which provides a Fourier expansion in the angular variable $\theta$:
\begin{equation}
    e^{i z \sin\theta} = \sum_{n=-\infty}^{\infty} J_n(z) e^{i n \theta}.
\end{equation}
The functions of different orders satisfy the following linear recurrence relations, which are highly useful for simplifying infinite series:
\begin{equation} \label{eq:bessel_recurrence}
    J_{n-1}(z) + J_{n+1}(z) = \frac{2n}{z} J_n(z), \qquad J_{n-1}(z) - J_{n+1}(z) = 2 J'_n(z).
\end{equation}
For Bessel functions of the first kind of integer order, the following summation identity holds for any $n\in\mathbb{Z}$:
\begin{equation} \label{eq:bessel_summation}
\sum_{\nu=-\infty}^{\infty}
J_\nu(z)J_{\nu+n}(z)
=
\delta_{n,0},
\end{equation}
where $\delta_{n,0}$ denotes the Kronecker delta.
In the small-argument limit ($z \to 0$), the lowest-order term dominates, yielding the asymptotic behaviors
\begin{equation} \label{eq:bessel_small_argument}
    J_0(z) = 1 - \frac{z^2}{4} + \mathcal{O}(z^4), \qquad J_{\pm 1}(z) = \pm \frac{z}{2} + \mathcal{O}(z^3),
\end{equation}
while $J_n(z) = \mathcal{O}(z^{|n|})$ for $|n| \ge 2$.

\subsection{Modified Bessel Functions of the Second Kind}
\label{app:bessel_second}

The modified Bessel function of the second kind, $K_n(x)$, is a linearly independent solution to the modified equation $x^2 y'' + x y' - (x^2 + n^2)y = 0$. For $\operatorname{Re}(x) > 0$, it is most conveniently evaluated using its integral representation:
\begin{equation}
    K_n(x) = \int_0^\infty e^{-x \cosh t} \cosh(nt) dt.
\end{equation}
An important related integral identity over the domain $(0, \infty)$, frequently encountered when integrating over the thermal distribution, is
\begin{equation}
    \int_0^\infty p^{2n} e^{-\frac{1}{a}\sqrt{p^2+m^2}} dp = (2n-1)!! m^{n+1} a^n K_{n+1}\left(\frac{m}{a}\right),
\end{equation}
where $n\in\mathbb{Z}_{\geq 0}$ and $a>0$. Unlike $J_n(z)$, $K_n(x)$ is even under reversal of its integer order, $K_{-n}(x) = K_n(x)$. Its recurrence relations and first derivative are given by
\begin{equation}
    K_{n+1}(x) - K_{n-1}(x) = \frac{2n}{x} K_n(x), \qquad K'_{n}(x) = -\frac{1}{2} \left[ K_{n-1}(x) + K_{n+1}(x) \right].
\end{equation}
For small arguments ($x \to 0$), the functions diverge:
\begin{equation}
    K_0(x) \approx -\ln\left(\frac{x}{2}\right) - \gamma, \qquad K_n(x) \approx \frac{(n-1)!}{2} \left(\frac{2}{x}\right)^n \quad (n \ge 1),
\end{equation}
where $\gamma$ is the Euler-Mascheroni constant. For large arguments ($x \to \infty$), the function decays exponentially, independent of the order $n$, yielding $K_n(x) \approx \sqrt{\pi/(2x)} \, e^{-x}$.

%%%%%%%%%%%%%%%%%%%%%%%%%%%%%%%%%%%%%%%%%%%%%%
\section{Global Equilibrium State and Gauge Invariance}
\label{app:equilibrium_gauge}
%%%%%%%%%%%%%%%%%%%%%%%%%%%%%%%%%%%%%%%%%%%%%%
In this appendix, we derive the global equilibrium state of a relativistic plasma in a general electromagnetic field within the framework of the relativistic Boltzmann equation (see, e.g., Ref.~\cite{groot1980relativistic}), explicitly demonstrate its gauge invariance, and subsequently apply it to the uniform magnetic field configuration utilized in our study.

\subsection{Global Equilibrium in a General Electromagnetic Field}
A necessary condition for a system to reach local equilibrium is that the entropy production vanishes, which requires the distribution function to satisfy the detailed balance condition in the collision integral. This restricts the logarithm of the distribution function to be a linear combination of the collisional invariants (particle number and four-momentum), yielding
\begin{equation}
\ln f_{\text{local}}(x, p) = a(x) + b_\mu(x) p^\mu.
\end{equation}
Thus, the local-equilibrium distribution function takes the following form
\begin{equation}
    f_{\text{local}}(x, p) =\exp\left[ a(x) + b_\mu(x) p^\mu\right],
    \label{local}
\end{equation}
where $a(x)$ and $b_\mu(x)$ are arbitrary spacetime-dependent functions.

For $f_{\text{local}}(x, p)$ to describe a global equilibrium state, it must also be an exact solution to the collisionless transport equation (i.e., the left-hand side of the Boltzmann equation must vanish). Substituting the form into this equation yields
\begin{equation}
p_\mu \left[ \partial^\mu a(x) + b_\nu(x) F^{\nu\mu}(x) \right] + p^\mu p^\nu \partial_\mu b_\nu(x) = 0.
\end{equation}
Since this equality must hold for arbitrary $p^\mu$, we can obtain two independent constraints:
\begin{align}
\partial^\mu b^\nu(x) + \partial^\nu b^\mu(x) &= 0, \label{eq:Killing}\\
\partial^\mu a(x) + b_\nu(x) F^{\nu\mu}(x) &= 0, \label{eq:a_constraint}
\end{align}
where the former equation is also called Killing condition. Excluding rigid rotations, its most general solution corresponds to a constant fluid four-velocity $u^\mu$ and a constant temperature $T$, allowing us to write $b^\mu = -u^\mu / T$. By defining the local chemical potential $\mu(x)$ such that $a(x) = \mu(x)/T$, Eq.~\eqref{eq:a_constraint} becomes
\begin{equation}
\partial^\mu \mu(x) - u_\sigma F^{\sigma\mu}(x) = 0.
\label{eq:mu_differential}
\end{equation}

Taking the spacetime derivative of Eq.~\eqref{eq:mu_differential} and extracting its antisymmetric part with respect to the indices $\mu$ and $\nu$, we obtain
\begin{equation}
u_\sigma \left[ \partial^\nu F^{\sigma\mu}(x) - \partial^\mu F^{\sigma\nu}(x) \right] = 0.
\end{equation}
With the help of the Bianchi identity, $\partial^\sigma F^{\mu\nu} + \partial^\mu F^{\nu\sigma} + \partial^\nu F^{\sigma\mu} = 0$, this relation simplifies to
\begin{equation}
u_\sigma \partial^\sigma F^{\mu\nu}(x) \equiv D F^{\mu\nu}(x) = 0,
\label{equilibrium condiction}
\end{equation}
where $D \equiv u_\sigma \partial^\sigma$ is the convective derivative along the fluid flow. Eq.~\eqref{equilibrium condiction} means that global equilibrium can only exist if the electromagnetic field tensor $F^{\mu\nu}(x)$ does not change in time in the global-rest frame as determined by $u^\mu$. The solution of Eq.~\eqref{equilibrium condiction} is
\begin{equation}
D A^\mu(x)  = \partial^\mu \Phi(x)
\label{eq:general_gauge}
\end{equation}
with $\Phi(x)$ an arbitrary function of the space-time coordinates. 
%By choosing a suitable gauge the function $\Phi(x)$ can be put equal to zero.

Substituting $F^{\sigma\mu}(x) = \partial^\sigma A^\mu(x) - \partial^\mu A^\sigma(x)$ back into Eq.~\eqref{eq:mu_differential}, we have
\begin{equation}
\partial^\mu \mu(x) - u_\sigma \partial^\sigma A^\mu(x) + u_\sigma \partial^\mu A^\sigma(x) = 0.
\label{equation for mu}
\end{equation}
Using the condition \eqref{eq:general_gauge}, Eq.~\eqref{equation for mu} becomes
\begin{equation}
\partial^\mu \left[ \mu(x) - \Phi(x) + u_\nu A^\nu(x) \right] = 0.
\end{equation}
Integrating this total gradient over spacetime yields 
\begin{equation}
\label{mu}
\mu(x) = \mu_0 + \Phi(x) - u_\nu A^\nu(x),
\end{equation}
where $\mu_0$ is an integration constant. Substituting $a(x) = \mu(x)/T$ and $b_\mu(x) = -u_\mu/T$ back into the local-equilibrium distribution function \eqref{local}, we obtain the general global equilibrium distribution function
\begin{equation}
f_{\text{global}}(x,p) = \exp \left\{ \frac{\mu_0 + \Phi(x) - \left[ p^\mu + A^\mu(x) \right] u_\mu}{T} \right\}.
\label{eq:general_eq_dist}
\end{equation}

\subsection{Gauge Independence of the Equilibrium Distribution}
It is crucial to verify that the global equilibrium distribution function \eqref{eq:general_eq_dist} is independent of the choice of gauge. Consider a $U(1)$ gauge transformation defined by an arbitrary scalar function $\Lambda(x)$:
\begin{equation}
A^\mu(x) \to A'^\mu(x) = A^\mu(x) + \partial^\mu \Lambda(x).
\end{equation}
Under this transformation, the electromagnetic field tensor $F^{\mu\nu}(x)$ remains invariant. The condition $u_\nu \partial^\nu A'^\mu = \partial^\mu \Phi'$ dictates how the scalar function $\Phi(x)$ must transform:
\begin{equation}
u_\nu \partial^\nu \left( A^\mu + \partial^\mu \Lambda \right) = \partial^\mu \Phi + \partial^\mu \left( u_\nu \partial^\nu \Lambda \right).
\end{equation}
Thus, $\Phi$ transforms as $\Phi'(x) = \Phi(x) + u_\nu \partial^\nu \Lambda(x)$.

Now, let us examine the combination in the exponent of Eq.~\eqref{eq:general_eq_dist}:
\begin{equation}
\mathcal{M}_{\text{gauge}}'(x) = \Phi'(x) - u_\mu A'^\mu(x) = \Phi(x) - u_\mu A^\mu(x) = \mathcal{M}_{\text{gauge}}(x).
\end{equation}
Consequently, the distribution function $f_{\text{global}}(x,p)$ is manifestly gauge-invariant. 

\subsection{Equilibrium State in a Uniform Magnetic Field}
We now apply this general result to the specific physical scenario investigated in our study: a plasma immersed in a uniform magnetic field $\mathbf{B}_0 = B_0 \hat{\mathbf{x}}$ with the fluid being macroscopically at rest, i.e., $u^\mu = u_0^\mu = (1, 0, 0, 0)$ and $T = T_0$.

For a stationary fluid, the relevant scalar function $\Phi(x)$ is determined by $u_{0\nu} \partial^\nu A^\mu = \partial_0 A^\mu = \partial^\mu \Phi$. Since a uniform and constant static magnetic field can always be described by a time-independent vector potential (for instance, in the Landau gauge $A^\mu = (0, 0, -B_0 z, 0)$), we have $\partial_0 A^\mu = 0$, which allows us to set $\Phi(x) = 0$.

Furthermore, since the fluid is at rest ($u_0^\mu = \delta^\mu_0$) and there is no external electric field ($A^0 = 0$), the gauge contraction term vanishes identically:
\begin{equation}
u_{0\mu} A^\mu(x) = A^0(x) = 0.
\end{equation}
The momentum contraction simply gives
\begin{equation}
u_{0\mu} p^\mu = p^0.
\end{equation}
Substituting the above results into Eq.~\eqref{eq:general_eq_dist}, the global equilibrium distribution function reduces to the standard, homogenous J\"{u}ttner form:
\begin{equation}
f_0(\mathbf{p}) = \exp\left( \frac{\mu_0 - p^0}{T_0} \right).
\end{equation}
%%%%%%%%%%%%%%%%%%%%%%%%%%%%%%%%%%%%%%%%%%%%%%
\section{Derivation of the Exact Series Solution}
\label{app:derivation}
%%%%%%%%%%%%%%%%%%%%%%%%%%%%%%%%%%%%%%%%%%%%%%
In this appendix, we provide a detailed, step-by-step derivation of the exact series solution presented in Eq.~\eqref{Exact Series Solution}, starting from the general solution in Eq.~\eqref{eq:general_solution_phi}.

First, let us examine the exponential integrating factors in Eq.~\eqref{eq:general_solution_phi} and Eq.~\eqref{eq:C_determined}. Using the explicit expression for $\mathcal{I}(\phi)$ defined in Eq.~\eqref{I}, we can expand the oscillatory part using the Jacobi--Anger identity $e^{iz\sin\theta} = \sum_{n=-\infty}^{\infty} J_n(z) e^{in\theta}$. Then we have
\begin{equation}
    e^{\pm \mathcal{I}(\phi)} = e^{\pm\left[ \frac{\alpha}{\omega_c}\phi+i\frac{p_\perp}{B_0}k_z\right]} e^{\pm i \xi \sin(\phi-\psi)} = e^{\pm\left[ \frac{\alpha}{\omega_c}\phi+i\frac{p_\perp}{B_0}k_z\right]} \sum_{n=-\infty}^{\infty} J_n(\xi) e^{\pm i n (\phi-\psi)}.
\end{equation}

Next, we define the integral $\mathcal{J}(\phi)$ as
\begin{align}
    \mathcal{J}(\phi) \equiv \int_{0}^{\phi} d\phi'\, e^{-\mathcal{I}(\phi')} \Delta(\phi') 
    &= \int_{0}^{\phi} d\phi'\, e^{-\frac{\alpha}{\omega_c}\phi'-i\frac{p_\perp}{B_0}k_z} \sum_{n=-\infty}^{\infty} J_n(\xi) e^{- i n (\phi'-\psi)} \sum_{m=-1}^{1} \Delta_m e^{im\phi'} \nonumber \\
    &= \sum_{n,m} J_n(\xi) \Delta_m e^{in\psi-i\frac{p_\perp}{B_0}k_z} \int_{0}^{\phi} d\phi'\, e^{\left[ -\frac{\alpha}{\omega_c} + i(m-n) \right] \phi'}.
\end{align}
Introducing the shorthand notation $\beta_{nm} \equiv -\frac{\alpha}{\omega_c} + i(m-n)$, the integration over $\phi'$ yields
\begin{equation}
    \mathcal{J}(\phi) = \sum_{n,m} J_n(\xi) \Delta_m e^{in\psi-i\frac{p_\perp}{B_0}k_z} \frac{e^{\beta_{nm}\phi} - 1}{\beta_{nm}}.
    \label{eq:app_J_phi}
\end{equation}

Now we evaluate the integration constant $C(p_x,p_\perp)$ using Eq.~\eqref{eq:C_determined}. Similarly, $\mathcal{J}(2\pi)$ becomes
\begin{equation}
    \mathcal{J}(2\pi) = \sum_{n,m} J_n(\xi) \Delta_m e^{in\psi-i\frac{p_\perp}{B_0}k_z} \frac{e^{-2\pi\alpha/\omega_c} - 1}{\beta_{nm}}.
\end{equation}
Substituting this into Eq.~\eqref{eq:C_determined}, we obtain the explicit form of $C(p_x,p_\perp)$:
\begin{align}
    C(p_x,p_\perp) &= \frac{1}{\omega_c} \frac{e^{2\pi\alpha/\omega_c}}{e^{2\pi\alpha/\omega_c}-1} \mathcal{J}(2\pi) \nonumber \\
    &= \frac{1}{\omega_c} \sum_{n,m} J_n(\xi) \Delta_m e^{in\psi-i\frac{p_\perp}{B_0}k_z} \left[ \frac{e^{2\pi\alpha/\omega_c}}{e^{2\pi\alpha/\omega_c}-1} \frac{e^{-2\pi\alpha/\omega_c} - 1}{\beta_{nm}} \right] \nonumber \\
    &= -\frac{1}{\omega_c} \sum_{n,m} J_n(\xi) \Delta_m e^{in\psi-i\frac{p_\perp}{B_0}k_z} \frac{1}{\beta_{nm}}.
    \label{eq:app_C_explicit}
\end{align}

Combining the integration constant in Eq.~\eqref{eq:app_C_explicit} with the integral term in Eq.~\eqref{eq:app_J_phi}, we can obtain the full distribution function $\delta\tilde{f}(p_x,p_\perp,\phi)$,
\begin{align}
    \delta \tilde f(p_x,p_\perp,\phi) 
    &= e^{\mathcal{I}(\phi)} \left[ C(p_x,p_\perp) - \frac{1}{\omega_c}\mathcal{J}(\phi) \right] \nonumber \\
    &= \left( e^{\frac{\alpha}{\omega_c}\phi} \sum_{l=-\infty}^{\infty} J_l(\xi) e^{il(\phi-\psi)} \right) \left( -\frac{1}{\omega_c} \sum_{n,m} J_n(\xi) \Delta_m e^{in\psi} \frac{e^{\left[-\frac{\alpha}{\omega_c} + i(m-n)\right]\phi}}{\beta_{nm}} \right) \nonumber \\
&= \sum_{l,n,m} \frac{\Delta_m J_n(\xi) J_l(\xi)}{-\beta_{nm} \omega_c} e^{i(l+m-n)\phi} e^{-i(l-n)\psi}.
\end{align}
Recalling the definition of $\beta_{nm}$, the denominator is $-\beta_{nm} \omega_c = -\left[ -\frac{\alpha}{\omega_c} + i(m-n) \right] \omega_c = \alpha + i(n-m)\omega_c$. This strictly recovers the exact series solution presented in Eq.~\eqref{Exact Series Solution}.

%%%%%%%%%%%%%%%%%%%%%%%%%%%%%%%%%%%%%%%%%%%%%%
\section{Explicit Components of the Retarded Current-Current Correlator}
\label{app:explicit_correlators}
%%%%%%%%%%%%%%%%%%%%%%%%%%%%%%%%%%%%%%%%%%%%%%
In this appendix we present the explicit expressions for all sixteen components of the retarded current-current correlator $G^{\mu\nu}_{JJ}(\omega,\mathbf k)$. These expressions underlie, and are algebraically equivalent to, the compact representation given in Eq.~\eqref{eq:compact_G}.
\begin{align}
    \begin{aligned} 
        G^{00}_{JJ}(\omega,\mathbf{k})=&(-i)\left(1-\sum\!\!\!\!\!\!\!\!\int \frac{ J_{l}(\xi) J_l(\xi)}{\frac{1}{\tau_R} - i(\omega - k_x v_x) + il\omega_c}\frac{f_0(\mathbf{p})}{T_0\tau_R\chi}\right)^{-1}\times\left(\sum\!\!\!\!\!\!\!\!\int \frac{J_{l}(\xi) J_l(\xi)}{\frac{1}{\tau_R} - i(\omega - k_x v_x) + il\omega_c}  \frac{f_0(\mathbf{p})}{T_0}v_xk_x \right.\\ &\left. +\sum\!\!\!\!\!\!\!\!\int \frac{J_{l+1}(\xi) J_l(\xi)}{\frac{1}{\tau_R} - i(\omega - k_x v_x) + il\omega_c}\frac{f_0(\mathbf{p})}{T_0} \frac{p_\perp k_{\perp}}{2p_0} +\sum\!\!\!\!\!\!\!\!\int \frac{J_{l-1}(\xi) J_l(\xi)}{\frac{1}{\tau_R} - i(\omega - k_x v_x) + il\omega_c}\frac{f_0(\mathbf{p})}{T_0} \frac{p_\perp k_{\perp}}{2p_0}\right), 
    \end{aligned}
\end{align}
\begin{align}
    \begin{aligned}
        G^{01}_{JJ}(\omega,\mathbf{k})=&(-i)\left(1-\sum\!\!\!\!\!\!\!\!\int \frac{ J_{l}(\xi) J_l(\xi)}{\frac{1}{\tau_R} - i(\omega - k_x v_x) + il\omega_c}\frac{f_0(\mathbf{p})}{T_0\tau_R\chi}\right)^{-1}\times\left(\sum\!\!\!\!\!\!\!\!\int \frac{J_{l}(\xi) J_l(\xi)}{\frac{1}{\tau_R} - i(\omega - k_x v_x) + il\omega_c}  \frac{f_0(\mathbf{p})}{T_0}v_x\omega \right),
    \end{aligned}
\end{align}
\begin{align}
    \begin{aligned}
        G^{02}_{JJ}(\omega,\mathbf{k})=&(-i)\left(1-\sum\!\!\!\!\!\!\!\!\int \frac{ J_{l}(\xi) J_l(\xi)}{\frac{1}{\tau_R} - i(\omega - k_x v_x) + il\omega_c}\frac{f_0(\mathbf{p})}{T_0\tau_R\chi}\right)^{-1}\\ &\times\left(\sum\!\!\!\!\!\!\!\!\int \frac{J_{l+1}(\xi) J_l(\xi)e^{i\psi}}{\frac{1}{\tau_R} - i(\omega - k_x v_x) + il\omega_c}\frac{f_0(\mathbf{p})}{T_0} \frac{p_\perp\omega}{2p_0} +\sum\!\!\!\!\!\!\!\!\int \frac{J_{l-1}(\xi) J_l(\xi)e^{-i\psi}}{\frac{1}{\tau_R} - i(\omega - k_x v_x) + il\omega_c}\frac{f_0(\mathbf{p})}{T_0} \frac{p_\perp\omega}{2p_0} \right),
    \end{aligned}
\end{align}
\begin{align}
    \begin{aligned}
        G^{03}_{JJ}(\omega,\mathbf{k})=&\left(1-\sum\!\!\!\!\!\!\!\!\int \frac{ J_{l}(\xi) J_l(\xi)}{\frac{1}{\tau_R} - i(\omega - k_x v_x) + il\omega_c}\frac{f_0(\mathbf{p})}{T_0\tau_R\chi}\right)^{-1}\\ &\times\left(-\sum\!\!\!\!\!\!\!\!\int \frac{J_{l+1}(\xi) J_l(\xi)e^{i\psi}}{\frac{1}{\tau_R} - i(\omega - k_x v_x) + il\omega_c}\frac{f_0(\mathbf{p})}{T_0} \frac{p_\perp\omega}{2p_0} +\sum\!\!\!\!\!\!\!\!\int \frac{J_{l-1}(\xi) J_l(\xi)e^{-i\psi}}{\frac{1}{\tau_R} - i(\omega - k_x v_x) + il\omega_c}\frac{f_0(\mathbf{p})}{T_0} \frac{p_\perp\omega}{2p_0} \right).
    \end{aligned}
\end{align}
\begin{align}
    \begin{aligned}
        G^{10}_{JJ}(\omega,\mathbf{k})=&G^{00}_{JJ}(\omega,\mathbf{k})\sum\!\!\!\!\!\!\!\!\int \frac{ J_{l}(\xi) J_l(\xi)}{\frac{1}{\tau_R} - i(\omega - k_x v_x) + il\omega_c}\frac{f_0(\mathbf{p})v_{x}}{T_0\tau_R\chi}-i\sum\!\!\!\!\!\!\!\!\int \frac{J_{l}(\xi) J_l(\xi)}{\frac{1}{\tau_R} - i(\omega - k_x v_x) + il\omega_c}  \frac{f_0(\mathbf{p})}{T_0}v_x^{2}k_x \\&-i\sum\!\!\!\!\!\!\!\!\int \frac{J_{l+1}(\xi) J_l(\xi)}{\frac{1}{\tau_R} - i(\omega - k_x v_x) + il\omega_c}\frac{f_0(\mathbf{p})}{T_0} \frac{p_\perp k_{\perp}v_x}{2p_0} -i\sum\!\!\!\!\!\!\!\!\int \frac{J_{l-1}(\xi) J_l(\xi)}{\frac{1}{\tau_R} - i(\omega - k_x v_x) + il\omega_c}\frac{f_0(\mathbf{p})}{T_0} \frac{p_\perp k_{\perp}v_x}{2p_0},
    \end{aligned}
\end{align}
\begin{align}
    \begin{aligned}
        G^{11}_{JJ}(\omega,\mathbf{k})=G^{01}_{JJ}(\omega,\mathbf{k})\sum\!\!\!\!\!\!\!\!\int \frac{ J_{l}(\xi) J_l(\xi)}{\frac{1}{\tau_R} - i(\omega - k_x v_x) + il\omega_c}\frac{f_0(\mathbf{p})v_{x}}{T_0\tau_R\chi}-i\sum\!\!\!\!\!\!\!\!\int \frac{ J_{l}(\xi) J_l(\xi)}{\frac{1}{\tau_R} - i(\omega - k_x v_x) + il\omega_c}\frac{f_0(\mathbf{p})v_{x}^2\omega}{T_0},
    \end{aligned}
\end{align}
\begin{align}
    \begin{aligned}
        G^{12}_{JJ}(\omega,\mathbf{k})=&G^{02}_{JJ}(\omega,\mathbf{k})\sum\!\!\!\!\!\!\!\!\int \frac{ J_{l}(\xi) J_l(\xi)}{\frac{1}{\tau_R} - i(\omega - k_x v_x) + il\omega_c}\frac{f_0(\mathbf{p})v_{x}}{T_0\tau_R\chi}-i\sum\!\!\!\!\!\!\!\!\int \frac{J_{l+1}(\xi) J_l(\xi)e^{i\psi}}{\frac{1}{\tau_R} - i(\omega - k_x v_x) + il\omega_c}\frac{f_0(\mathbf{p})}{T_0} \frac{p_\perp v_{x}\omega}{2p_0} \\& -i\sum\!\!\!\!\!\!\!\!\int \frac{J_{l-1}(\xi) J_l(\xi)e^{-i\psi}}{\frac{1}{\tau_R} - i(\omega - k_x v_x) + il\omega_c}\frac{f_0(\mathbf{p})}{T_0} \frac{p_\perp v_{x}\omega}{2p_0},
    \end{aligned}
\end{align}
\begin{align}
    \begin{aligned}
        G^{13}_{JJ}(\omega,\mathbf{k})=&G^{03}_{JJ}(\omega,\mathbf{k})\sum\!\!\!\!\!\!\!\!\int \frac{ J_{l}(\xi) J_l(\xi)}{\frac{1}{\tau_R} - i(\omega - k_x v_x) + il\omega_c}\frac{f_0(\mathbf{p})v_{x}}{T_0\tau_R\chi}-\sum\!\!\!\!\!\!\!\!\int \frac{J_{l+1}(\xi) J_l(\xi)e^{i\psi}}{\frac{1}{\tau_R} - i(\omega - k_x v_x) + il\omega_c}\frac{f_0(\mathbf{p})}{T_0} \frac{p_\perp v_{x}\omega}{2p_0} \\&+\sum\!\!\!\!\!\!\!\!\int \frac{J_{l-1}(\xi) J_l(\xi)e^{-i\psi}}{\frac{1}{\tau_R} - i(\omega - k_x v_x) + il\omega_c}\frac{f_0(\mathbf{p})}{T_0} \frac{p_\perp v_{x}\omega}{2p_0},
    \end{aligned}
\end{align}
\begin{align}
    \begin{aligned}
        G^{20}_{JJ}(\omega,\mathbf{k})=&G^{00}_{JJ}(\omega,\mathbf{k})\sum\!\!\!\!\!\!\!\!\int\sum_{s=\pm1} \frac{ J_{l+s}(\xi) J_l(\xi)e^{is\psi}}{\frac{1}{\tau_R} - i(\omega - k_x v_x) + i(l+s)\omega_c}\frac{p_\perp}{2p_0}\frac{f_0(\mathbf{p})}{T_0\tau_R\chi}\\&-i\sum\!\!\!\!\!\!\!\!\int\sum_{s=\pm1} \frac{ J_{l+s}(\xi) J_l(\xi)e^{is\psi}}{\frac{1}{\tau_R} - i(\omega - k_x v_x) + i(l+s)\omega_c}\frac{p_\perp v_x k_x}{2p_0}\frac{f_0(\mathbf{p})}{T_0}
        \\&-i\sum\!\!\!\!\!\!\!\!\int\sum_{s=\pm1} \frac{ J_{l+s+1}(\xi) J_l(\xi)e^{i(s+1)\psi}}{\frac{1}{\tau_R} - i(\omega - k_x v_x) + i(l+s)\omega_c}\frac{f_0(\mathbf{p})}{T_0}\left(\frac{p_\perp}{2p_0}\right)^2\left(k_y-ik_z\right)
        \\&-i\sum\!\!\!\!\!\!\!\!\int\sum_{s=\pm1} \frac{ J_{l+s-1}(\xi) J_l(\xi)e^{i(s-1)\psi}}{\frac{1}{\tau_R} - i(\omega - k_x v_x) + i(l+s)\omega_c}\frac{f_0(\mathbf{p})}{T_0}\left(\frac{p_\perp}{2p_0}\right)^2\left(k_y+ik_z\right),
    \end{aligned}
\end{align}
\begin{align}
    \begin{aligned}
        G^{21}_{JJ}(\omega,\mathbf{k})=&G^{01}_{JJ}(\omega,\mathbf{k})\sum\!\!\!\!\!\!\!\!\int\sum_{s=\pm1} \frac{ J_{l+s}(\xi) J_l(\xi)e^{is\psi}}{\frac{1}{\tau_R} - i(\omega - k_x v_x) + i(l+s)\omega_c}\frac{p_\perp}{2p_0}\frac{f_0(\mathbf{p})}{T_0\tau_R\chi}\\&-i\sum\!\!\!\!\!\!\!\!\int\sum_{s=\pm1} \frac{ J_{l+s}(\xi) J_l(\xi)e^{is\psi}}{\frac{1}{\tau_R} - i(\omega - k_x v_x) + i(l+s)\omega_c}\frac{p_\perp v_x \omega}{2p_0}\frac{f_0(\mathbf{p})}{T_0},
    \end{aligned}
\end{align}
\begin{align}
    \begin{aligned}
        G^{22}_{JJ}(\omega,\mathbf{k})=&G^{02}_{JJ}(\omega,\mathbf{k})\sum\!\!\!\!\!\!\!\!\int\sum_{s=\pm1} \frac{ J_{l+s}(\xi) J_l(\xi)e^{is\psi}}{\frac{1}{\tau_R} - i(\omega - k_x v_x) + i(l+s)\omega_c}\frac{p_\perp}{2p_0}\frac{f_0(\mathbf{p})}{T_0\tau_R\chi}\\&-i\sum\!\!\!\!\!\!\!\!\int\sum_{s=\pm1} \frac{ J_{l+s+1}(\xi) J_l(\xi)e^{i(s+1)\psi}}{\frac{1}{\tau_R} - i(\omega - k_x v_x) + i(l+s)\omega_c}\frac{f_0(\mathbf{p})}{T_0}\left(\frac{p_\perp}{2p_0}\right)^2\omega
        \\&-i\sum\!\!\!\!\!\!\!\!\int\sum_{s=\pm1} \frac{ J_{l+s-1}(\xi) J_l(\xi)e^{i(s-1)\psi}}{\frac{1}{\tau_R} - i(\omega - k_x v_x) + i(l+s)\omega_c}\frac{f_0(\mathbf{p})}{T_0}\left(\frac{p_\perp}{2p_0}\right)^2\omega,
    \end{aligned}
\end{align}
\begin{align}
    \begin{aligned}
        G^{23}_{JJ}(\omega,\mathbf{k})=&G^{03}_{JJ}(\omega,\mathbf{k})\sum\!\!\!\!\!\!\!\!\int\sum_{s=\pm1} \frac{ J_{l+s}(\xi) J_l(\xi)e^{is\psi}}{\frac{1}{\tau_R} - i(\omega - k_x v_x) + i(l+s)\omega_c}\frac{p_\perp}{2p_0}\frac{f_0(\mathbf{p})}{T_0\tau_R\chi}\\&-\sum\!\!\!\!\!\!\!\!\int\sum_{s=\pm1} \frac{ J_{l+s+1}(\xi) J_l(\xi)e^{i(s+1)\psi}}{\frac{1}{\tau_R} - i(\omega - k_x v_x) + i(l+s)\omega_c}\frac{f_0(\mathbf{p})}{T_0}\left(\frac{p_\perp}{2p_0}\right)^2\omega
        \\&+\sum\!\!\!\!\!\!\!\!\int\sum_{s=\pm1} \frac{ J_{l+s-1}(\xi) J_l(\xi)e^{i(s-1)\psi}}{\frac{1}{\tau_R} - i(\omega - k_x v_x) + i(l+s)\omega_c}\frac{f_0(\mathbf{p})}{T_0}\left(\frac{p_\perp}{2p_0}\right)^2\omega,
    \end{aligned}
\end{align}
\begin{align}
    \begin{aligned}
        G^{30}_{JJ}(\omega,\mathbf{k})=&G^{00}_{JJ}(\omega,\mathbf{k})\sum\!\!\!\!\!\!\!\!\int\sum_{s=\pm1}\left(-is\right) \frac{ J_{l+s}(\xi) J_l(\xi)e^{is\psi}}{\frac{1}{\tau_R} - i(\omega - k_x v_x) + i(l+s)\omega_c}\frac{p_\perp}{2p_0}\frac{f_0(\mathbf{p})}{T_0\tau_R\chi}\\&-\sum\!\!\!\!\!\!\!\!\int\sum_{s=\pm1} s\frac{ J_{l+s}(\xi) J_l(\xi)e^{is\psi}}{\frac{1}{\tau_R} - i(\omega - k_x v_x) + i(l+s)\omega_c}\frac{p_\perp v_x k_x}{2p_0}\frac{f_0(\mathbf{p})}{T_0}
        \\&-\sum\!\!\!\!\!\!\!\!\int\sum_{s=\pm1} s\frac{ J_{l+s+1}(\xi) J_l(\xi)e^{i(s+1)\psi}}{\frac{1}{\tau_R} - i(\omega - k_x v_x) + i(l+s)\omega_c}\frac{f_0(\mathbf{p})}{T_0}\left(\frac{p_\perp}{2p_0}\right)^2\left(k_y-ik_z\right)
        \\&-\sum\!\!\!\!\!\!\!\!\int\sum_{s=\pm1} s\frac{ J_{l+s-1}(\xi) J_l(\xi)e^{i(s-1)\psi}}{\frac{1}{\tau_R} - i(\omega - k_x v_x) + i(l+s)\omega_c}\frac{f_0(\mathbf{p})}{T_0}\left(\frac{p_\perp}{2p_0}\right)^2\left(k_y+ik_z\right),
    \end{aligned}
\end{align}
\begin{align}
    \begin{aligned}
        G^{31}_{JJ}(\omega,\mathbf{k})=&G^{01}_{JJ}(\omega,\mathbf{k})\sum\!\!\!\!\!\!\!\!\int\sum_{s=\pm1}(-is) \frac{ J_{l+s}(\xi) J_l(\xi)e^{is\psi}}{\frac{1}{\tau_R} - i(\omega - k_x v_x) + i(l+s)\omega_c}\frac{p_\perp}{2p_0}\frac{f_0(\mathbf{p})}{T_0\tau_R\chi}\\&-\sum\!\!\!\!\!\!\!\!\int\sum_{s=\pm1} s\frac{ J_{l+s}(\xi) J_l(\xi)e^{is\psi}}{\frac{1}{\tau_R} - i(\omega - k_x v_x) + i(l+s)\omega_c}\frac{p_\perp v_x \omega}{2p_0}\frac{f_0(\mathbf{p})}{T_0},
    \end{aligned}
\end{align}
\begin{align}
    \begin{aligned}
        G^{32}_{JJ}(\omega,\mathbf{k})=&G^{02}_{JJ}(\omega,\mathbf{k})\sum\!\!\!\!\!\!\!\!\int\sum_{s=\pm1} (-is)\frac{ J_{l+s}(\xi) J_l(\xi)e^{is\psi}}{\frac{1}{\tau_R} - i(\omega - k_x v_x) + i(l+s)\omega_c}\frac{p_\perp}{2p_0}\frac{f_0(\mathbf{p})}{T_0\tau_R\chi}\\&-\sum\!\!\!\!\!\!\!\!\int\sum_{s=\pm1} s\frac{ J_{l+s+1}(\xi) J_l(\xi)e^{i(s+1)\psi}}{\frac{1}{\tau_R} - i(\omega - k_x v_x) + i(l+s)\omega_c}\frac{f_0(\mathbf{p})}{T_0}\left(\frac{p_\perp}{2p_0}\right)^2\omega
        \\&-\sum\!\!\!\!\!\!\!\!\int\sum_{s=\pm1} s\frac{ J_{l+s-1}(\xi) J_l(\xi)e^{i(s-1)\psi}}{\frac{1}{\tau_R} - i(\omega - k_x v_x) + i(l+s)\omega_c}\frac{f_0(\mathbf{p})}{T_0}\left(\frac{p_\perp}{2p_0}\right)^2\omega,
    \end{aligned}
\end{align}
\begin{align}
    \begin{aligned}
        G^{33}_{JJ}(\omega,\mathbf{k})=&G^{03}_{JJ}(\omega,\mathbf{k})\sum\!\!\!\!\!\!\!\!\int\sum_{s=\pm1} (-is)\frac{ J_{l+s}(\xi) J_l(\xi)e^{is\psi}}{\frac{1}{\tau_R} - i(\omega - k_x v_x) + i(l+s)\omega_c}\frac{p_\perp}{2p_0}\frac{f_0(\mathbf{p})}{T_0\tau_R\chi}\\&+i\sum\!\!\!\!\!\!\!\!\int\sum_{s=\pm1}s \frac{ J_{l+s+1}(\xi) J_l(\xi)e^{i(s+1)\psi}}{\frac{1}{\tau_R} - i(\omega - k_x v_x) + i(l+s)\omega_c}\frac{f_0(\mathbf{p})}{T_0}\left(\frac{p_\perp}{2p_0}\right)^2\omega
        \\&-i\sum\!\!\!\!\!\!\!\!\int\sum_{s=\pm1} s\frac{ J_{l+s-1}(\xi) J_l(\xi)e^{i(s-1)\psi}}{\frac{1}{\tau_R} - i(\omega - k_x v_x) + i(l+s)\omega_c}\frac{f_0(\mathbf{p})}{T_0}\left(\frac{p_\perp}{2p_0}\right)^2\omega.
    \end{aligned}
\end{align}

\bibliographystyle{JHEP}
\bibliography{ref.bib}

\end{document}